\documentclass[conference,a4paper,twocolumn]{IEEEtran}
\usepackage{amsmath}
\usepackage{amssymb}
\usepackage[dvips]{graphicx}

\begin{document}

\sloppy

\title{Single-Letter Characterization of Epsilon-Capacity for Mixed Memoryless Channels}

\author{
  \IEEEauthorblockN{Hideki Yagi}
  \IEEEauthorblockA{Dept.\ of Communication Engineering \& Informatics\\
The University of Electro-Communications \\
Chofu-shi, Tokyo 182-8585, Japan\\
Email: h.yagi@uec.ac.jp} 
  \and
  \IEEEauthorblockN{Ryo Nomura}
  \IEEEauthorblockA{School of Network and Information\\
    Senshu University\\
    Kawasaki-shi, Kanagawa 214-8580, Japan\\
    Email: nomu@isc.senshu-u.ac.jp}
}


\def\bf#1{\boldsymbol{#1}}
\def\vect#1{\boldsymbol{#1}}
\def\QED{\hfill$\Box$}

\newcommand{\Tr}{\textrm{Tr}}
\newcommand{\E}{\mathsf{E}}	
\newcommand{\V}{\mathsf{V}}	
\newcommand{\ep}{\textrm{ep}}
\newcommand{\cl}{\textrm{cl}}
\newcommand{\Cov}{\textrm{Cov}}		
\newcommand{\Qinv}{Q_\textrm{inv}}	
\newcommand{\Ginv}{G_\textrm{inv}}
\newcommand{\Code}{\mathcal{C}_n}
\newcommand{\PX}{\mathcal{P}(\mathcal{X})}			
\def\bf#1{\boldsymbol{#1}}
\def\vect#1{\boldsymbol{#1}}
\def\QED{\hfill$\Box$}

\newtheorem{e_theo}{Theorem}
\newtheorem{e_defin}{Definition}
\newtheorem{e_lem}{Lemma}
\newtheorem{e_prop}{Proposition}
\newtheorem{e_assump}{Assumption}
\newtheorem{e_coro}{Corollary}
\newtheorem{e_rema}{Remark}
\newtheorem{e_exam}{Example}

\renewcommand{\baselinestretch}{0.99} 
\setlength{\abovedisplayskip}{4.0pt} 
\setlength{\belowdisplayskip}{4.0pt} 
\setlength{\jot}{4.0pt} 

\maketitle

\begin{abstract}
 For the class of mixed channels decomposed into stationary memoryless channels, single-letter characterizations of the $\varepsilon$-capacity have not been known except for restricted classes of channels such as the regular decomposable channel introduced by Winkelbauer.
This paper gives single-letter characterizations of $\varepsilon$-capacity for mixed channels decomposed into at most countably many memoryless channels with a finite input alphabet and a general output alphabet with/without cost constraints.
It is shown that a given characterization reduces to the one for the channel capacity given by Ahlswede when $\varepsilon$ is zero. 
In the proof of the coding theorem, the meta converse bound, originally given by Polyanskiy, Poor and Verd\'u, is particularized for the mixed channel decomposed into general component channels.
\end{abstract}

\section{Introduction}

The maximum rate of sequence of codes that can attain a decoding error probability less than $\varepsilon \in [0,1)$ is called the \emph{$\varepsilon$-capacity}.
It is well-known that stationary memoryless channels have the so-called \emph{strong converse property}, and the $\varepsilon$-capacity coincides with the channel capacity ($\varepsilon$-capacity with $\varepsilon=0$) \cite{Wolfowitz78}.
On the other hand, allowing a decoding error probability up to $\varepsilon$, the maximum achievable rate may be improved for non-stationary and/or non-ergodic channels.
The simplest example is \emph{mixed channels} \cite{Han2003} (also referred to as decomposable channels \cite{Winkelbauer71a} or averaged channels \cite{Ahlswede68,Kieffer2007}) whose probability distribution is characterized by a mixture of multiple stationary memoryless channels.
This channel is stationary but non-ergodic, and is theoretically important as basic example to be investigated when extensions of coding theorems for ergodic channels are addressed. 
This channel is known to give the simplest mathematical model of (non-ergodic) block fading channels (c.f.\ \cite{Telatar99,YDKP2013}).

For general channels including mixed channels, a general formula of $\varepsilon$-capacity has been given by Verd\'u and Han \cite{Verdu-Han94}.
This formula, however, involves limit operations with respect to the code length $n$, and thus is infeasible to calculate in general.
On the other hand, for mixed channels decomposed into stationary memoryless channels with a finite input alphabet, a single-letter characterization of the channel capacity has been given by Ahlswede \cite{Ahlswede68}.
This characterization is of importance because the channel capacity can be computed with the complexity independent of $n$. 
However, to the best of authors' knowledge,  no single-letter characterizations of the $\varepsilon$-capacity have been known, or at least no rigorous proofs of an expression have  appeared in the literature. 
The regular decomposable channel which is decomposed into memoryless channels, introduced by Winkelbauer \cite{Winkelbauer71a}, is an example of channel classes for which a single-letter characterization of $\varepsilon$-capacity has been given.

This paper gives a single-letter characterization of the $\varepsilon$-capacity for mixed channels decomposed into stationary memoryless channels with a finite input alphabet and a general output alphabet.
First, a single-letter characterization of the $\varepsilon$-capacity is given for mixed channels decomposed into at most countably many stationary memoryless channels\footnote{A single-letter expression of the capacity has also been given by Ahlswede \cite{Ahlswede68} for the mixed channel averaged by an arbitrary probability measure, and the expression has been simplified by Han \cite{Han2003}. 
Other related studies which analyze the maximum rate  for which the outage probability is admitted up to $\varepsilon$ for a non-ergodic block fading channel has been given by \cite{Telatar99} and \cite{YDKP2013}.}.
An alternative expression is also provided, and it is shown that the characterization reduces to the one for the channel capacity given by Ahlswede \cite{Ahlswede68} when $\varepsilon$ is zero.
Then the theorem is extended to the case when input symbols are subject to a cost constraint.
The coding theorems are proved by the \emph{information spectrum method} (c.f.\ \cite{Han2003,Verdu-Han94}) combined with recently developed analytical methods for the finite blocklength regime (e.g., \cite{Hayashi2009,PPV2010,Tomamichel-Tan2013a,VTGM2013}).
In the proof of the coding theorems, the so-called \emph{meta converse} bound \cite{PPV2010}, which is known as the best converse bound to date is particularized for mixed channels\footnote{Although the meta converse bound also applies to mixed channels, it should be modified to finely analyze fundamental limits of codes.}.
With this bound, kinds of previously known converse bounds developed for general channels may also be particularized for the mixed channel setting.

\section{Preliminaries}

\subsection{General Channel and $\varepsilon$-Capacity}

Consider a channel $W^n: \mathcal{X}^n \rightarrow \mathcal{Y}^n$ which stochastically  maps an input sequence $X^n \in \mathcal{X}^n$ of length $n$ into an output sequence $Y^n \in \mathcal{Y}^n$. 
Here,  $\mathcal{X}$ and $\mathcal{Y}$ denote a finite input alphabet and an arbitrary output alphabet\footnote{In the case where $\mathcal{Y}$ is abstract in general, we understand that $W^n(\vect{y}|\vect{x})$ and $P_{Y^n}(\vect{y})$ denote the corresponding probability measures $W^n(d\vect{y}|\vect{x})$ and $P_{Y^n}(d\vect{y})$, respectively, and that $\log\frac{W^n(\vect{y}|\vect{x})}{P_{Y^n}(\vect{y})}$ denotes the Radon-Nikodym derivative $\log\frac{W^n(d \vect{y}|\vect{x})}{P_{Y^n}(d \vect{y})}$. As in \cite{Han2003}, we keep the notation simple and use the summation $\sum$ to denote the integral $\int$, too.}, respectively. 
We denote by $\mathcal{P}(\mathcal{X})$ the set of all probability mass functions on $\mathcal{X}$. 
A sequence $\vect{W} := \{ W^n\}_{n=1}^\infty$ of channels $W^n$ is referred to as a \emph{general channel} \cite{Han2003}.


Let $\Code$ be a code of length $n$ and the number of codewords $|\Code| = M_n$ with an encoding function $\phi: \{1,\ldots, M_n\} \rightarrow \mathcal{X}^n$ and a decoding function $\psi : \mathcal{Y}^n \rightarrow \{1,\ldots, M_n\}$.


\begin{e_defin}{\rm
The \emph{average}  probability of decoding error over $W^n$ is defined as
\begin{eqnarray}
{P_e}(\Code) := \frac{1}{M_n}\sum_{i=1}^{M_n} \Pr[\psi(Y^n) \neq i |\, i ~\mathrm{sent}]. 
\end{eqnarray}
The code $\Code$ is referred to as an $(n,M_n, {P_e}(\Code) )$ code.
\QED}
\end{e_defin}
\begin{e_rema}
The \emph{maximum} error probability defined as
\begin{align}
e(\Code) := \max_{i \in \{1,\cdots,M_n \}}  \Pr[\psi(Y^n) \neq i | i ~\textrm{sent}]
\end{align}
has also been considered in the literature. 
All the capacity results in this paper are also valid under the maximum error probability criterion.
\QED
\end{e_rema}


\begin{e_defin}\label{def:1st_achievable}
{\rm
A coding rate $R \ge 0$ is said to be \emph{achievable} if there exists a sequence of $(n,M_n, P_e(\Code) )$ codes satisfying 
\begin{align}
\limsup_{n \rightarrow \infty} {P_e}(\Code) \le \varepsilon ~~~\mathrm{and}~~~\liminf_{n \rightarrow \infty} \frac{1}{n} \log M_n \ge R. \label{eq:error_achievable} 
\end{align}
The supremum of  $\varepsilon$-achievable rates is called the $\varepsilon$-\emph{capacity} and is denoted by $C(\varepsilon|\vect{W})$.
}\QED
\end{e_defin}

\begin{e_rema} \label{rema:right_continuous}
The $\varepsilon$-capacity $C(\varepsilon|\vect{W})$ is a right-continuous function in $\varepsilon$ \cite{Verdu-Han94}. 
\QED
\end{e_rema}

\begin{e_rema} \label{rema:different_def}
{\rm
An $\varepsilon$-achievable rate is often defined 
by replacing \eqref{eq:error_achievable} with 
\begin{align}
&{P_e}(\Code) \le \varepsilon ~~\mathrm{and}~~ \frac{1}{n} \log M_n \ge R - \lambda \label{eq:conv2}
\end{align}
(e.g., \cite{Kieffer2007,Verdu-Han94,Winkelbauer71a}, etc.).  
The $\varepsilon$-capacity in this case is not right-continuous in $\varepsilon$, and  the provided characterizations of the $\varepsilon$-capacity are valid except at most countably many discontinuous points of $\varepsilon$-capacity (c.f.\ \cite[Theorem 6]{Verdu-Han94}). 
}
\QED
\end{e_rema}

\subsection{Mixed Memoryless Channel} \label{sect:Mixed_memoryless_channel}

Consider a set of at most countably many $\vect{W}_\ell := \{W_\ell^n\}_{n=1}^\infty~(\ell = 1,2,\cdots)$, and the set of indices of $\vect{W}_\ell$ is denoted by $\Omega$. 
The \emph{mixed channel} decomposed into $\{\vect{W}_\ell \}_{\ell \in \Omega}$ is defined by
\begin{align}
W^n(\vect{y}|\vect{x}) = \sum_{\ell \in \Omega} w_\ell W_\ell^n(\vect{y}|\vect{x}),~~(\forall \vect{x} \in \mathcal{X}^n, \forall  \vect{y} \in \mathcal{Y}^n), \label{eq:mixed_channel}
\end{align}
a mixture of $\{W_\ell^n\}$  with the mixing ratio $\{ w_\ell \ge 0 \}_{\ell=1}^\infty$ satisfying $\sum_{\ell=1}^\infty w_\ell = \sum_{\ell \in \Omega} w_\ell =1$.
{Hereafter, we assume that $w_\ell > 0$ for all $\ell \in \Omega$, for simplicity.} 
Each $\vect{W}_\ell$ is called a \emph{component channel} or simply components. 
Given an input probability distribution $P_{X^n}$, the output from $W_\ell^n$ induced by the input $X^n$ is denoted by $Y_\ell^n$. That is,
\begin{align}
P_{X^nY_\ell^n}(\vect{x},\vect{y}) = P_{X^n}(\vect{x}) W_\ell^n(\vect{y}|\vect{x}) ~~ (\forall \vect{x} \in \mathcal{X}^n, \forall \vect{y} \in \mathcal{Y}^n). \nonumber
\end{align}

The mixed channel $\vect{W}$ given by at most countably many stationary memoryless channels $\{W_\ell \}_{\ell \in \Omega}$ satisfying
$W_\ell^n (\vect{y}|\vect{x}) = \prod_{i=1}^n W_\ell(y_i|x_i)$
is called the \textbf{mixed memoryless channel}.
{Hereafter, we assume that the input alphabet $\mathcal{X}$ is finite and the output alphabet $\mathcal{Y}$ may be infinite as long as 
the mutual information $I_{P_X}(X;Y_\ell)$ calculated by $P_X$ and $W_\ell$ is continuous in $P_X$ for all $\ell \in \Omega$.
For example, if $\mathcal{Y}$ is a complete separable metric space, then $I_{P_X}(X;Y_\ell)$ is concave and continuous in $P_X$ \cite[Lemma 3]{Csiszar92}.}

\section{Main Theorems}

\subsection{General Mixed Memoryless Channels}

The following theorem gives a single-letter characterization of the $\varepsilon$-capacity.
\begin{e_theo} \label{theo:mixed_DMC_e-cap}
Let $\vect{W}$ be a mixed memoryless channel with $|\mathcal{X}| < \infty$.
For any fixed $\varepsilon \in [0,1)$, the $\varepsilon$-capacity is given by
\begin{align}
C(\varepsilon | \vect{W}) =  \sup_{ P_X \in \PX } \sup \left\{ R \, \Big| F_w(R|P_X) \le \varepsilon  \right\}, \label{eq:e-Cap}
\end{align}
where
\begin{align}
F_w(R|P_X) :=   \sum_{\ell \in \Omega} w_\ell \vect{1} \left\{ I_{P_X} (X;Y_\ell) \le R \right\}.    
\end{align}
Here, $I_{P_X}(X;Y_\ell)$ denotes the mutual information calculated by $P_X$ and $W_\ell$, and $\vect{1}\{ A \}$ denotes the indicator function which takes one if a proposition $A$ is true and takes zero otherwise.
\end{e_theo}
(Proof)~~A proof is given in Sect.\ \ref{sect:proof_e-Cap}.
\QED



We define the function $A: \mathcal{P}(\mathcal{X}) \times [0,1] \rightarrow \mathbb{R}$ as
\begin{align}
A(P_X, \delta) := \sup \left\{ R~|~ F_w(R|P_X) \le \delta\right\}, \label{eq:func_A}
\end{align}
where $\mathbb{R}$ denotes the set of real numbers.
The $\varepsilon$-capacity given by Theorem \ref{theo:mixed_DMC_e-cap} is expressed as 
\begin{align}
C(\varepsilon|\vect{W}) = \sup_{P_X \in \mathcal{P}(\mathcal{X})} A(P_X, \varepsilon). \label{eq:mixed_DMC_e_cap2}
\end{align}
%
Let $\mathcal{D}$ be a compact set in $\mathcal{P}(\mathcal{X})$. Some properties of the function $A(P_X, \delta)$ and $\tilde{R}(\delta | \mathcal{D}):=\sup_{P_X \in \mathcal{D}} A(P_X, \delta)$ are shown by the following lemma. 
\begin{e_lem} \label{lem:property_A}
For the functions $A(P_X,\delta)$ and $\tilde{R}(\delta | \mathcal{D})$, the following hold:
\begin{itemize}
\item[(a)] $A(P_X,\delta)$ is continuous in $P_X$ for fixed $\delta$.
\item[(b)] $A(P_X,\delta)$ is non-decreasing in $\delta$ for fixed $P_X$.
\item[(c)] $A(P_X,\delta)$ is right-continuous in $\delta$. That is, $\lim_{\delta \downarrow \delta_0} A(P_X, \delta) = A(P_X, \delta_0)$.
\item[(d)] $\tilde{R}(\delta | \mathcal{D})$ is right-continuous in $\delta$.
\end{itemize}
\end{e_lem}
(Proof)~~
Properties (b) and (c) are easily verified by the definition of $A(P_X, \delta)$.
Proofs of Properties (a) and (d) are given in Appendix \ref{sec:convexity_A2} and Appendix \ref{sec:prop_tilde_R}, respectively.
\QED

The function $F_w(R|P_X)$ which appears in the definition of $A(P_X, \delta)$ is not continuous in $P_X$ obviously. 
It is of interest to see that the function $A(P_X, \delta)$ has Property (a) nevertheless. 
By Property (a), there exists at least one $P_X \in \mathcal{P}(\mathcal{X})$ that gives $\tilde{R}(\delta | \mathcal{D})$.
That is,  $\tilde{R}(\delta | \mathcal{D})= \max_{P_X \in \mathcal{D}} A(P_X, \delta) $.
The first supremum on the right-hand side (r.h.s.) of \eqref{eq:e-Cap} is actually maximum. 

By Properties (b) and (d), the $\varepsilon$-capacity given in Theorem \ref{theo:mixed_DMC_e-cap} can also be expressed as 
 \begin{align}
C(\varepsilon|\vect{W}) = \inf_{\delta>0} \sup_{P_X \in \mathcal{P}(\mathcal{X})} \sup \left\{ R~|~ F_w(R|P_X) \le \varepsilon + \delta\right\} \label{eq:mixed_DMC_e_cap3}
\end{align}
To prove Theorem \ref{theo:mixed_DMC_e-cap},  it is sufficient to show that \eqref{eq:mixed_DMC_e_cap3} holds, and this fact is used in Sect.\ \ref{sect:proof_e-Cap}.


\subsection{Special Case: Well-Ordered Mixed Memoryless Channels}

It is shown that the single-letter characterization in Theorem \ref{theo:mixed_DMC_e-cap} reduces a previously known expression for a restricted class of mixed channels.
As an example, the following class of mixed memoryless channels is introduced.

Let $C_\ell$ denote the channel capacity of the $\ell$-th component channel $W_\ell$ and $\Pi_{\ell}$ be the set of input probability distributions that achieve $C_\ell$.
Without loss of generality, we assume that the component channels are indexed to satisfy $C_\ell \le C_{\ell+1}$, where components $W_i$ and $W_j~(i \neq j)$ such that $C_i=C_j$ are arbitrarily indexed if $|\Omega| < \infty$.
\begin{e_defin}[Well-Ordered Mixed Memoryless Channel]
For each $\ell \in \Omega$, if there exists some $P_X \in \Pi_\ell$ such that
\begin{align}
C_\ell \le I_{P_X}(X;Y_j) ~~\mbox{for~all}~~j: C_\ell \le C_j,  \label{eq:C_assumption1}
\end{align}
then the mixed channel $\vect{W}$ is  said to be \textbf{well-ordered}.
\QED
\end{e_defin}

For example, let us consider {a well-ordered} mixed memoryless channel of two components $\vect{W}_1 = \{W_1^n\}_{n=1}^\infty, \vect{W}_2 = \{W_2^n\}_{n=1}^\infty$.  By the condition \eqref{eq:C_assumption1}, it should hold $C_1 \le C_2$ and
\begin{align}
& C_1  \le I(X;Y_2) ~~(\exists P_X \in \Pi_1).  \label{eq:C_assumption2}
\end{align}
When $C_1 = C_2$, \eqref{eq:C_assumption1} requires 
$\Pi_1 \cap \Pi_2 \neq \emptyset$.
If the component channels $\{W_\ell | \ell \in \Omega \}$ are all output-symmetric (e.g., the mixed BSCs \cite{Kieffer2007}), then the condition \eqref{eq:C_assumption1} trivially holds. 

It it readily shown that {every well-ordered mixed memoryless channel} is an instance of \emph{regular decomposable} channels introduced by  Winkelbauer \cite{Winkelbauer71a}. 
The $\varepsilon$-capacity of a regular decomposable channel has been given by \cite{Winkelbauer71a}.
For well-ordered mixed memoryless channels, the following corollary follows from Theorem \ref{theo:mixed_DMC_e-cap}.
\begin{e_coro}[Winkelbauer \cite{Winkelbauer71a}] \label{coro:mixed_DM_RDC_e-cap}
Let $\vect{W}$ be a well-ordered mixed memoryless channels such that $|\mathcal{X}| < \infty$, and define
\begin{align}
\tilde{F}_w(R) :=   \sum_{\ell \in \Omega} w_\ell \vect{1} \left\{ C_\ell \le R \right\}.    
\end{align}
For any $\varepsilon \in [0,1)$, the $\varepsilon$-capacity is given by
\begin{align}
C(\varepsilon | \vect{W}) = \sup \left\{ R \, \Big|  \tilde{F}_w(R)  \le \varepsilon \right\}, \label{eq:epsilon_cap}
\end{align}
\end{e_coro}
\QED

Corollary \ref{coro:mixed_DM_RDC_e-cap} slightly extends the coding theorem by Winkelbauer \cite{Winkelbauer71a}  for the well-ordered mixed memoryless channel to the case of non-discrete $\mathcal{Y}$.

{Consider the case $|\Omega| < \infty$. By \eqref{eq:epsilon_cap}, the  $\varepsilon$-capacity of the mixed channel satisfying \eqref{eq:C_assumption1} is given by $C(\varepsilon|\vect{W}) = C_{k^*}$, where $k^*$ is the component index satisfying 
\begin{align}
\sum_{\ell \in \Omega} w_\ell \vect{1} \{ C_\ell < C_{k^*} \} \le \varepsilon < \tilde{F}_w (C_{k^*}). \label{eq:epsilon_cap2}
\end{align}
For example, the $\varepsilon$-capacity for the well-ordered mixed channel $\vect{W}$ with $|\Omega| = 3$ is given by
\begin{align}
C(\varepsilon|\vect{W}) = \left\{
\begin{array}{cl}
C_{1}, & \mathrm{if}~\varepsilon \in [0,w_1) \\
C_{2}, & \mathrm{if}~\varepsilon \in [w_1,w_1+w_2) \\
C_{3}, & \mathrm{otherwise} 
\end{array}
\right. .\label{eq:C_assumption4}
\end{align}
}

It is of interest to see that the expression of the $\varepsilon$-capacity in Corollary \ref{coro:mixed_DM_RDC_e-cap} 
is similar to the one for the channel with states \cite{Tomamichel-Tan2014}. Specifically, Example 1 in
\cite{Tomamichel-Tan2014} deals with the mixed channel decomposable into finitely many (not necessarily well-ordered) stationary memoryless components, and
both the encoder and the decoder can access the channel state
information, which corresponding to the index of component
channels in this paper. In this case, the expression of the $\varepsilon$-capacity coincides with the one given in \eqref{eq:epsilon_cap}.
This fact implies that the optimum rate without the channel state information is the same as the one with the channel state information if the mixed channel is well-ordered.

\subsection{Alternative Expression of $\varepsilon$-Capacity}

We give an alternative expression of the $\varepsilon$-capacity of the mixed memoryless channel given by Theorem \ref{theo:mixed_DMC_e-cap}.
We first show the following lemma.

\begin{e_lem} \label{lem:alternative_express}
Let $\vect{W}$ be a mixed memoryless channel with $|\mathcal{X}| < \infty$. Then, we have
\begin{align}
\sup\left\{ R | \, F_w(R|P_X) \le \varepsilon \right\} &= \sup_{\{S \subseteq \Omega | \, w(S) \ge 1 -\varepsilon\}} \inf_{\ell \in S} I_{P_X}(X;Y_\ell) \label{eq:alternative_express}
\end{align}
for all $P_X \in \mathcal{P}(\mathcal{X})$, where $w(S)$ denotes $\sum_{\ell \in S} w_\ell$.
\end{e_lem}
(Proof)~~See Appendix \ref{sec:proof_alternative_express}.
\QED


Combining \eqref{eq:e-Cap} with Lemma \ref{lem:alternative_express} provides an alternative expression of the $\varepsilon$-capacity as
\begin{align}
C(\varepsilon|\vect{W}) =\sup_{P_X} \sup_{\{S \subseteq \Omega | \, w(S) \ge 1 -\varepsilon \}} \inf_{\ell \in S} I_{P_X}(X;Y_\ell) \label{eq:alternative_e-cap2}
\end{align}
in the case of at most countably many component channels.
When $\varepsilon = 0$, the r.h.s.\ of \eqref{eq:alternative_e-cap2} becomes $\sup_{P_X}  \inf_{\ell \in \Omega} I_{P_X}(X;Y_\ell)$, which coincides with the capacity expression given by Ahlswede \cite{Ahlswede68}. 


{On the r.h.s.\ of \eqref{eq:alternative_e-cap2}, $\inf_{\ell \in S} I_{P_X}(X;Y_\ell)$ with $w(S) \ge 1-\varepsilon$ is the infimum of concave functions of $P_X$.
When $\varepsilon = 0$, $\inf_{\ell \in S} I_{P_X}(X;Y_\ell) = \inf_{\ell \in \Omega} I_{P_X}(X;Y_\ell)$ is concave  in $P_X$. 
When $\varepsilon > 0$, however, this function  is not necessarily concave since the domain $S$ with $w(S) \ge 1-\varepsilon$ depends on $P_X$. 
}  

Similar to \eqref{eq:alternative_e-cap2}, the $\varepsilon$-capacity of a well-ordered mixed memoryless channel can also be expressed as
\begin{align}
C(\varepsilon|\vect{W}) = \sup_{\{S \subseteq \Omega | \, w(S) \ge 1 -\varepsilon \}} \inf_{\ell \in S} \, C_\ell. \label{eq:alternative_e-cap_well_oredered}
\end{align}

\subsection{$\varepsilon$-Capacity under Cost Constraint} \label{sect:cost_constraint}

We now turn to considering the coding for which an input symbol $X$ is constrained by a cost function $c:\mathcal{X} \rightarrow \mathbb{R}$.
This problem includes the power constraint over the channel with a continuous alphabet such as the additive white Gaussian noise (AWGN) channel as an instance.

If every codeword $\phi(i)~(\forall i \in \{1,\ldots, M_n\})$ of a code $\Code$ is restricted to be in the set
\begin{align}
\mathcal{X}_{c,\Gamma} := \left\{ \vect{x} \in \mathcal{X}^n | \, \sum_{i=1}^{n} c(x_i) \le n \Gamma \right\},
\end{align}
this condition is referred to as the \emph{cost constraint $\Gamma$}. 
A code $\Code$ attains an error probability $\varepsilon \in [0,1)$ under the cost constraint $\Gamma$ is called an \emph{$(n,M_n, \varepsilon, \Gamma)$ code}.  

\begin{e_defin}
If \eqref{eq:error_achievable} holds under the cost constraint $\Gamma$, then the rate $R$ is said to be $(\varepsilon,\Gamma)$-achievable. 
The supremum of $(\varepsilon,\Gamma)$-achievable rates for  $\vect{W}$ is referred to as the \emph{$(\varepsilon,\Gamma)$-capacity} and is denoted by $C(\varepsilon, \Gamma | \vect{W})$.
\QED
\end{e_defin}

The following theorem characterizes the optimum coding rate under a cost constraint for the mixed memoryless channel. 
\begin{e_theo} \label{theo:mixed_DMC_e-Gamma-cap}
Let $\vect{W}$ be a mixed memoryless channel with $|\mathcal{X}| < \infty$. The $(\varepsilon, \Gamma)$-capacity for a given $\Gamma \in \mathbb{R}$ and $\varepsilon \in [0,1)$ is given by
\begin{align}
C(\varepsilon, \Gamma | \vect{W}) = \! \! \sup_{ P_X : \E_{P_X} c(X) \le \Gamma } \!  \sup \left\{ R \, \Big|  F_w(R|P_X)  \le \varepsilon  \right\}. \label{eq:e-Gamma-Cap}
\end{align}
\end{e_theo}
(Proof)~~Converse Part is exactly the same line as the one for Theorem \ref{theo:mixed_DMC_e-cap}.
To prove Direct Part, we use an ensemble of constant composition codes whose type $P_0$ satisfies the constraint $\Gamma$ and $M_n$ codeword are chosen by the uniform distribution on the set of sequences with type $P_0$. 
We can apply an information spectrum approach by Hayashi \cite[Sect.\ X-B]{Hayashi2009} to the proof of Direct Part of Theorem \ref{theo:mixed_DMC_e-cap}, showing that any rate $R$ less than the r.h.s.\ of \eqref{eq:e-Gamma-Cap} is $(\varepsilon, \Gamma)$-achievable. 
\QED

The set of $P_X \in \mathcal{P}(\mathcal{X})$ such that $\E_{P_X} c(X) \le \Gamma$ is closed convex, and hence is compact. Then from Property (a) in Lemma \ref{lem:property_A}, the first supremum in \eqref{eq:e-Gamma-Cap} is maximum, and  from Property (d), the r.h.s\ in \eqref{eq:e-Gamma-Cap} is right-continuous in $\varepsilon$.
When $\varepsilon=0$, \eqref{eq:e-Gamma-Cap} reduces to the capacity under a cost constraint $\Gamma$:
\begin{align}
C(0,\Gamma|\vect{W}) = \sup_{ P_X : \E_{P_X} c(X) \le \Gamma } \inf_{\ell \in \Omega} I_{P_X}(X;Y_\ell),
\end{align}
{which has been shown by Han \cite{Han2003}.}

{The function $C(\varepsilon, \Gamma | \vect{W}) $ is referred to as the \emph{capacity-cost function}, which is analogous to the rate-distortion function for lossy source coding (c.f.\ \cite{Han2003}).
The capacity-cost function is also referred to as the \emph{capacity-expense function}, and some of its properties for discrete memoryless channels (DMCs) have been shown in \cite{Blahut72}.
By definition, the capacity-cost function is monotonic nondecreasing in $\Gamma$. We show some properties of the capacity-cost function.
\begin{e_theo}
The capacity-cost function has the following properties:
\begin{itemize}
\item[(i)] concave in $\Gamma$ for $\Gamma > 0$;
\item[(ii)] strictly increasing in $\Gamma$ for $0 \le \Gamma < \Gamma^*$, where $\Gamma^*$ is the minimum cost for which the capacity cost-function coincides with the $\varepsilon$-capacity;
\item[(iii)] if $\Gamma < \Gamma^*$, then $C(\varepsilon,\Gamma|\vect{W})$ is achieved by some $P_X \in \mathcal{P}(\mathcal{X})$ such that $\E_{P_X} c(X) = \Gamma$.
\QED
\end{itemize}
\end{e_theo}
These properties, which can be shown in an analogous way to the proofs in \cite[Appendix]{Blahut72}, are handed down from the capacity-cost function for DMCs.
However, unlike the DMC case, the set of optimum input distributions that achieve the $\varepsilon$-capacity under a cost constraint is not necessarily convex.
}

\section{One-Shot Error Bounds for Mixed Channel}

The proof of Theorem \ref{theo:mixed_DMC_e-cap} provided in Sect.\ \ref{sect:proof_e-Cap} uses so-called ``one-shot" error bounds which hold for the mixed channel decomposed into (not necessarily stationary or ergodic) general component channels. 
 
 
First we show converse (lower) error bounds. Following \cite[Sect.\ III-D]{PPV2010}, we introduce simple hypothesis testing: Given an observation $Z \in \mathcal{Z}$ according to either of two probability measures $P, Q$ on $\mathcal{Z}$,  consider a hypothesis test 
\begin{align}
H_0: Z \sim P ~~\mbox{vs.}~~~ H_1: Z \sim Q
\end{align}
to judge the true probability measure.
When observing $Z$, a test $\xi: \mathcal{Z} \rightarrow \{0,1\}$ judges $P$ to be true with probability $\xi(Z)$ and $Q$ to be true with probability $1-\xi(Z)$. 
The error event  when the true measure is $P$ is called \emph{the error of the first kind} and the one when the true measure is $Q$ is called \emph{the error of the second kind}.
For a fixed $\alpha \in [0,1]$, the optimum test that minimizes the error probability of the second kind among those whose error probability of the first kind satisfies  
$\sum_{z \in \mathcal{Z}} P(z) (1-\xi(z)) \le  \alpha$ is denoted by $\xi^*$, and its error probability of the second kind is denoted by
\begin{align}
\beta_{\alpha}(P, Q) &:=\min_{\substack{\xi:\mathcal{Z} \rightarrow \{0,1\} : \\ \sum_{z} P(z)(1-\xi(z)) \le \alpha}} \sum_{z \in \mathcal{Z}} Q(z) \xi(z).
\end{align}
Likewise, let $\alpha_{\beta}(P, Q) $ denote the  minimum error probability of the second kind among tests whose error probability of the first kind is less than or equal to $\beta$.

The following lemma particularizes a meta converse bound by Polyanskiy, Poor, and Verd\'u \cite{PPV2010} for the mixed channels. 

\begin{e_lem}[Meta Converse for Mixed Channel] \label{lem:mixed_meta_conv}
Let $\{ Q_{Y_\ell^n} \}_{\ell \in \Omega}$ be a set of arbitrary probability measures.
Then every $(n, M_n, \varepsilon_n)$ code $\Code$ with a (possibly probabilistic) decoding function $\xi: \mathcal{Y}^n \rightarrow \{1,2,\ldots, M_n\}$ satisfies 
\begin{align}
\varepsilon_n &\ge \sum_{\ell \in \Omega} w_\ell \, \alpha_{\frac{1}{M_n}} (P_{X^n} W_\ell^n, P_{X^n} Q_{Y_\ell^n}) \label{eq:mixed_meta_conv_alpha}
\end{align}
and
\begin{align}
\frac{1}{M_n} \ge \sum_{\ell \in \Omega} w_\ell \beta_{\varepsilon_n^{(\ell)}}(P_{X^n}W_\ell^n, P_{X^n}Q_{Y_\ell^n}). \label{eq:mixed_meta_conv_beta}
\end{align}
Here, $P_{X^n}$ is the uniform distribution on $\Code$, and  {$\varepsilon_n^{(\ell)}$ denotes the average probability of decoding error over $W_\ell^n$ given by  
\begin{align}	
& \hspace*{-1.5mm}\varepsilon_n^{(\ell)} := 1 - \frac{1}{M_n}\sum_{i=1}^{M_n} \sum_{\vect{y} \in \mathcal{Y}^n} W_\ell^n(\vect{y}|\phi(i)) \xi(i|\vect{y}) ~~(\forall \ell \in \Omega), \label{eq:epsilon_n^ell}
\end{align}
where $\phi(i)$ denotes the codeword assigned to message $i$, and $ \xi(i|\vect{y}) $  denotes the probability of $i$ being estimated given $\vect{y}$.}
\end{e_lem}
(Proof)~~The first inequality is due to \cite{VTGM2013}, and the second one is due to \cite{PPV2010}. A proof is given in Appendix \ref{sec:lem_proof2}. 
\QED

The following lemma is established by modifying a lemma shown by Tomamichel and Tan \cite{Tomamichel-Tan2013a} for mixed channels. 
\begin{e_lem} \label{lem:relaxed_meta_conv}
Given a family of pairs of probability measures $\{(P_\ell, Q_\ell)  \}_{\ell \in \Omega}$ on $\mathcal{Z}$, consider a hypothesis test
\begin{align}
H_0: Z_\ell \sim P_\ell ~~\mbox{vs.}~~~ H_1: Z_\ell \sim Q_\ell  \label{eq:HT_ell}
\end{align}
for each $\ell \in \Omega$. 
For any given $\varepsilon \in [0,1)$, letting $\{\varepsilon_{\ell} \in [0,1)\}_{\ell \in \Omega}$ be a sequence such that $\sum_{\ell \in \Omega} w_\ell \varepsilon_{\ell} = \varepsilon$, we have 
\begin{align}
- \log \sum_{\ell \in \Omega} w_\ell \beta_{\varepsilon_{\ell}}(P_\ell, Q_\ell) \le D_{{\rm s}}^{\varepsilon + \delta}(\{P_\ell\} || \{ Q_\ell\}) - \log \delta  \label{eq:mixed_meta_conv2} 
\end{align}
with an arbitrary constant $\delta \in (0,1] $, where $D_{{\rm s}}^{\varepsilon} (\{P_\ell\}|| \{ Q_\ell\})$ denotes the value
\begin{align}
\sup \left\{ R \, \Big|~ \sum_{\ell \in \Omega} w_\ell P_\ell \left\{ \log \frac{P_\ell(Z_\ell) }{Q_\ell(Z_\ell)} \le R \right\} \le \varepsilon \right\}.
\end{align}
\end{e_lem}
(Proof)~~A proof is given in Appendix \ref{sec:lem_proof3}. 
\QED


We set $P_\ell := P_{X^n} \times W_\ell^n, Q_\ell := P_{X^n} \times Q_{Y_\ell^n} $, $\varepsilon := \varepsilon_n$, and $\varepsilon_\ell := \varepsilon_n^{(\ell)}$ in Lemma \ref{lem:relaxed_meta_conv}.
Since $\varepsilon_n^{(\ell)}$ given in \eqref{eq:epsilon_n^ell} satisfies $\sum_{\ell \in \Omega} w_\ell \varepsilon_n^{(\ell)} = \varepsilon_n$, \eqref{eq:mixed_meta_conv2} holds. 
Then from \eqref{eq:mixed_meta_conv_beta}, every $(n,M_n, \varepsilon_n)$ code $\Code$ satisfies \begin{align}
\log M_n \le   D_{\rm s}^{\varepsilon_n + \delta} (\{P_{X^n}W_\ell^n\}|| \{ P_{X^n}Q_{Y_\ell^n}\}) - \log \delta \label{eq:mixed_meta_conv2b}
\end{align}
with an arbitrary constant $\delta \in (0,1]$.  

{
\begin{e_rema}
It is easily verified that Lemmas \ref{lem:mixed_meta_conv} and \ref{lem:relaxed_meta_conv} can be extended to the mixed channel with a general mixture (c.f.\ \cite[Sect.\ 3.3]{Han2003}).
In this case, the summand should be replaced with integral.
\QED
\end{e_rema}
} 
 

We next consider upper (achievability) error bounds. 
The following lemma particularizes 
 the Feinstein upper bound \cite{Feinstein54} for the mixed channels. 
\begin{e_lem} \label{lem:Feinstein_mixed_channel}
For any given $P_{X^n} \in \mathcal{P}(\mathcal{X}^n)$, there exists an $(n,M_n,\varepsilon_n)$ code satisfying 
\begin{align}
& \varepsilon_n \le \sum_{\ell \in \Omega} w_\ell \Pr\left\{ \frac{1}{n} \log \frac{W_\ell^n(Y_\ell^n|X^n)}{P_{Y_\ell^n}(Y_\ell^n)} \le \frac{1}{n} \log M_n \right. \nonumber \\
&~~~~~~~~~~~~~~~~~~~~~~\left. + \, \gamma + \frac{1}{n} \log \frac{1}{w_\ell}\right\} + e^{-n\gamma}, \label{eq:mixed_Feinstein_bound}
\end{align}
where $\gamma > 0$ is an arbitrary constant and  $P_{Y^n}$ denotes the marginal measure $P_{Y^n}(\vect{y}) = \sum_{\vect{x} \in \mathcal{X}^n} P_{X^n}(\vect{x}) W_\ell^n (\vect{y}|\vect{x})$.
\QED
\end{e_lem}

Equation \eqref{eq:mixed_Feinstein_bound} can be derived by the result shown by Han \cite[Lemma 1.4.1]{Han2003}. 
Although the original bound by Han uses a sequence $\{\gamma_n \ge 0 | \, \lim_{n \rightarrow \infty} \gamma_n = 0 \}$ instead of a constant $\gamma$, 
an examination verifies that \eqref{eq:mixed_Feinstein_bound} holds for any constant $\gamma>0$.

\section{Proof of Theorem \ref{theo:mixed_DMC_e-cap}} \label{sect:proof_e-Cap}

\subsection{Converse Part} \label{sect:proof_e-Cap_converse}


For a given $\vect{x} \in \mathcal{X}^n$, we denote 
$W_{\ell|\vect{x}}^n := W_\ell^n(\cdot| \vect{x})$ for simplicity.
For  a given $P_X \in \PX$, we define
\begin{align}
(P_X W_\ell)^{\times n} (\vect{y}):=\prod_{i=1}^n \sum_{x \in \mathcal{X}} P_X(x) W_\ell(y_i | x).
\end{align}

Converse Part of Theorem \ref{theo:mixed_DMC_e-cap} is stated as follows: 


\begin{e_theo}[Converse Theorem] \label{theo:mixed_DMC_e-cap_conv}
For a mixed channel $\vect{W}$, any $\varepsilon$-achievable rate $R$ for $\varepsilon \in [0,1)$ satisfies\begin{align}
R \le  {\inf_{\delta > 0}} \sup_{ P_X \in \PX } \sup \left\{ R \, \Big|  F_w(R|P_X)  \le \varepsilon {+ \delta} \right\}. \label{eq:converse_theorem}
\end{align}
\QED
\end{e_theo}

Before stating the proof of Converse Part, we give some preliminaries.
By the Chebyshev inequality, the following lemma holds:
\begin{e_lem} \label{lem:Chebyshev_ineq}
For any fixed $\vect{x} \in \mathcal{X}^n$, we denote its type (empirical distribution) by $P_n$. Let $\gamma > 0$ be an arbitrary constant and define 
\begin{align}
& \mathcal{B}_{\ell|\vect{x}}^{(n)} (\gamma) := \left\{ \vect{y}\, \Big| \left| \frac{1}{n} \log \frac{W_\ell^n(\vect{y}|\vect{x})}{(P_n W_\ell)^{\times n}(\vect{y})} -I_{P_n}(X;Y_\ell)\right| \le \gamma \right\}\label{eq:set_B_l_x}
 \end{align}
for all $\ell \in \Omega$.
Then we have 
 \begin{align}
 W_{\ell|\vect{x}}^n \left\{ Y_\ell^n \in  \mathcal{B}_{\ell|\vect{x}}^{(n)} (\gamma) \right\} \ge 1 - \frac{A(\gamma)}{n} \label{eq:Chebyshev_ineq}
 \end{align}
with a constant $A(\gamma) \ge 0$ independent of $n$, $P_n$ and $\ell$. 
\QED
\end{e_lem}
The conditional variance of information density $\log \frac{W_\ell(Y_\ell|X)}{(P_X W_\ell)(Y_\ell)}$ given $P_X$,
\begin{align}
V(P_X,W_\ell) :=  \E_{P_X} \left[\V_{W_\ell}\left[\log \frac{W_\ell(Y_\ell|X)}{(P_X W_\ell)(Y_\ell)}  \Big | X  \right] \right], 
\end{align}
is upper bounded by
$\V_{P_XW_\ell}\left[ \log \frac{W_\ell(Y_\ell|X)}{(P_X W_\ell)(Y_\ell)} \right]$, which can be verified as follows  (see also \cite[Lemma 62]{PPV2010}): defining
\begin{align}
U_1&:= \E \left[ \E\left[ \log \frac{W_\ell(Y_\ell|X)}{(P_X W_\ell)(Y_\ell)} \Big| X \right]^2  \right], \nonumber \\
U_2&:= \E \left[ \log \frac{W_\ell(Y_\ell|X)}{(P_X W_\ell)(Y_\ell)}  \right]^2, 
\end{align}
then $\E\left[ \log \frac{W_\ell(Y_\ell|X)}{(P_X W_\ell)(Y_\ell)} \Big| X \right]^2$ is a convex function of $P_X$ since $f(z) := z^2$ is convex and nondecreasing for $z \ge 0$, and $g(x) := \E\left[ \log \frac{W_\ell(Y_\ell|x)}{(P_X W_\ell)(Y_\ell)} \right]$ is convex.
Therefore, we obtain $U_1 \ge U_2$, which leads to the claim.
The variance $\V_{P_XW_\ell}\left[ \log \frac{W_\ell(Y_\ell|X)}{(P_X W_\ell)(Y_\ell)} \right]$ is further bounded uniformly by $\frac{8|\mathcal{X}|}{e^2}$ \cite[Remark 3.1.1]{Han2003}, the constant $A(\gamma)$  in \eqref{eq:Chebyshev_ineq} can be chosen independently of $\ell\in \Omega$ and  $P_n \in \mathcal{T}_n$.


We are now in a position to prove Theorem \ref{theo:mixed_DMC_e-cap_conv}.
Let $R$ be $\varepsilon$-achievable.
Then, from \eqref{eq:error_achievable}, there exists a sequence of $(n, M_n, \varepsilon_n)$ codes $\Code$ with some $\{ \delta_n \ge 0 | \delta_1 \ge \delta_2 \ge \cdots \ge 0, \lim_{n \rightarrow \infty} \delta_n = 0\}$ satisfying  
\begin{align}
\frac{1}{n} \log M_n &\ge R - \gamma  ~\mathrm{and}~ \varepsilon_n \le \varepsilon + \delta_n  ~(\exists n_1 > 0; \forall n \ge n_1)  \label{eq:e-achievable_rate} 
\end{align}
for an arbitrarily fixed constant $\gamma > 0$. 
%
Borrowing an idea given by Hayashi \cite[Sect. X-A]{Hayashi2009}, we set $\delta = \frac{1}{n}$ and  
\begin{align}
Q_{Y_\ell^n} (\vect{y}) = \frac{1}{|\mathcal{T}_n|} \sum_{P_n \in \mathcal{T}_n} (P_n W_\ell)^{\times n} (\vect{y}) ~~~(\forall \vect{y} \in \mathcal{Y}^n)\label{eq:Q_Y_ell}
\end{align}
in \eqref{eq:mixed_meta_conv2b}, where $\mathcal{T}_n$ denotes the set of types on $\mathcal{X}^n$. 
We define 
\begin{align}
R_n^* := \frac{1}{n} D_{{\rm s}}^{\varepsilon_n  + \frac{1}{n}}(\{P_{X^n}W_\ell^n\}|| \{ P_{X^n}Q_{Y_\ell^n}\}) + \frac{1}{n} \log n.  \label{eq:R_n^*}
\end{align}
Since
the first term on the r.h.s.\ is expressed as
\begin{align}
\sup \left\{ R \, \Big| \sum_{\ell \in \Omega} w_\ell \Pr\left\{ \frac{1}{n} \log \frac{W_\ell^n(Y_\ell^n|X^n) }{Q_{Y_\ell^n}(Y_\ell^n)} {\le} R \right\} \le \varepsilon_n {+} \frac{1}{n} \right\}, \nonumber
\end{align}
it can be verified that  
there exists an $\vect{x}_0 \in \Code$ such that  
\begin{align}
& \sum_{\ell \in \Omega} w_\ell W_{\ell|\vect{x}_0}^n \left\{ \frac{1}{n}  \log \frac{W_\ell^n(Y_\ell^n|\vect{x}_0) }{Q_{Y_\ell^n}(Y_\ell^n)} \le R_n^*  - \frac{1}{n} \log n -\gamma \right\} \nonumber \\
&~~~~~\le \varepsilon_n + \frac{1}{n}  \label{eq:particular_x}
\end{align}
as follows: By definition in \eqref{eq:R_n^*}, we can re-express 
\begin{align}
&R_n^* - \frac{1}{n} \log n \nonumber \\
&= \sup \left\{ R \, \Big| \sum_{\ell \in \Omega} w_\ell \Pr\left\{ \frac{1}{n} \log \frac{W_\ell^n(Y_\ell^n|X^n) }{Q_{Y_\ell^n}(Y_\ell^n)} {\le} R \right\} \le \varepsilon_n {+} \frac{1}{n} \right\}  \nonumber \\
 &= \sup \left\{ R \, \Big| \! \sum_{\vect{x} \in \Code} \frac{1}{M_n} \sum_{\ell \in \Omega} w_\ell W_{\ell|\vect{x}}^n \left\{ \frac{1}{n}  \log \frac{W_\ell^n(Y_\ell^n|\vect{x}) }{Q_{Y_\ell^n}(Y_\ell^n)} \le R \right\} \right. \nonumber \\
&~~~~~~~~~~~~~~~~\left. \le \varepsilon_n + \frac{1}{n} \right\}.  \label{eq:R_n^*_ineq}
\end{align}
Suppose that \eqref{eq:particular_x} does not hold for any $\vect{x} \in \Code$. Then we have
\begin{align}
\frac{1}{M_n}\sum_{\vect{x}  \in \Code} &   \sum_{\ell \in \Omega} w_\ell W_{\ell|\vect{x}_0}^n \left\{ \frac{1}{n}  \log \frac{W_\ell^n(Y_\ell^n|\vect{x}_0) }{Q_{Y_\ell^n}(Y_\ell^n)} \right. \nonumber \\
& \left. \le R_n^*  - \frac{1}{n} \log n - \gamma \right\} > \varepsilon_n + \frac{1}{n}, \label{eq:particular_x2}
\end{align}
and this implies that $R_n^*  - \frac{1}{n} \log n - \gamma$ is strictly greater than the r.h.s. of \eqref{eq:R_n^*_ineq}.
Since this contradicts \eqref{eq:R_n^*_ineq}, it is concluded that there exists at least one $\vect{x}_0 \in \Code$ satisfying \eqref{eq:particular_x}.  
Denoting by $P_0^n$ the type of $\vect{x}_0$, we have a chain of inequalities
\begin{align}
&W_{\ell|\vect{x}_0}^n \left\{ \frac{1}{n}  \log \frac{W_\ell^n(Y_\ell^n|\vect{x}_0) }{Q_{Y_\ell^n}(Y_\ell^n)} \le R_n^*  - \frac{1}{n} \log n -\gamma \right\} \nonumber \\
 &\ge W_{\ell|\vect{x}_0}^n \left\{\frac{1}{n}  \log \frac{W_\ell^n(Y_\ell^n|\vect{x}_0) }{(P_0^n W_\ell)^{\times n}(Y_\ell^n)} \le R_n^* - \frac{1}{n} \log n|\mathcal{T}_n| -\gamma \right\} \nonumber \\
&\ge W_{\ell|\vect{x}_0}^n \left\{\frac{1}{n}  \log \frac{W_\ell^n(Y_\ell^n|\vect{x}_0) }{(P_0^n W_\ell)^{\times n}(Y_\ell^n)} \le R_n^* - \frac{1}{n} \log n|\mathcal{T}_n| -\gamma, \right. \nonumber \\
&~~~~~~~~~~~~~~~~\left. Y_\ell^n \in \mathcal{B}_{\ell|\vect{x}_0}^{(n)}(\gamma) \right\} \nonumber \\
& \ge \vect{1} \left\{  I_{P_0^n} (X;Y_\ell) \le R_n^* {-} \frac{1}{n} \log n|\mathcal{T}_n| - 2 \gamma  \right\} - \frac{A(\gamma)}{n},  \label{eq:particular_x3} 
\end{align}
where $\mathcal{B}_{\ell|\vect{x}_0}^{(n)}(\gamma) $ is defined in \eqref{eq:set_B_l_x} and $A(\gamma) \ge  0$ is a constant independent of $n$, $P_0^n$, and $\ell$.
We use the relation in \eqref{eq:Q_Y_ell} for the first inequality.
The inequality in \eqref{eq:particular_x3} can be verified since (i)  for $\ell \in \Omega$ such that $I_{P_0^n} (X;Y_\ell)  \le R_n^*  - \frac{1}{n} \log n|\mathcal{T}_n| - 2 \gamma$, we have 
\begin{align}
& W_{\ell|\vect{x}_0}^n \left\{\frac{1}{n}  \log \frac{W_\ell^n(Y_\ell^n|\vect{x}_0) }{(P_0^n W_\ell)^{\times n}(Y_\ell^n)} \le R_n^* - \frac{1}{n} \log n|\mathcal{T}_n| -\gamma, \right. \nonumber \\
&~~~~~~~~~~\left. Y_\ell^n \in \mathcal{B}_{\ell|\vect{x}_0}^{(n)}(\gamma) \right\} \nonumber \\
&~~= W_{\ell|\vect{x}_0}^n \left\{Y_\ell^n \in \mathcal{B}_{\ell|\vect{x}_0}^{(n)}(\gamma) \right\} \ge 1- \frac{A(\gamma)}{n}
\end{align}
by Lemma \ref{lem:Chebyshev_ineq} and (ii) for $\ell \in \Omega$ such that $I_{P_0^n} (X;Y_\ell) > R_n^*  - \frac{1}{n} \log n|\mathcal{T}_n| -2 \gamma$, a trivial lower bound
\begin{align}
& W_{\ell|\vect{x}_0}^n \left\{\frac{1}{n}  \log \frac{W_\ell^n(Y_\ell^n|\vect{x}_0) }{(P_0^n W_\ell)^{\times n}(Y_\ell^n)} \le R_n^* - \frac{1}{n} \log n|\mathcal{T}_n| -\gamma, \right. \nonumber \\
&~~~~~~~~~~\left.Y_\ell^n \in \mathcal{B}_{\ell|\vect{x}_0}^{(n)}(\gamma) \right\} \ge - \frac{A(\gamma)}{n}
\end{align}
holds.
Note that the r.h.s.\ of \eqref{eq:particular_x3} depends on $P_0^n \in \mathcal{T}_n$ but not on individual codewords. 
Since $A(\gamma) \ge 0$ is a constant independent of $\ell$ and $P_0^n$,
we obtain
\begin{align}
&\sum_{\ell} w_\ell  \vect{1} \left\{  I_{P_0^n} (X;Y_\ell) \le R_n^*  - \frac{1}{n} \log n|\mathcal{T}_n| - 2 \gamma  \right\}  \nonumber \\
 &~~~~~ \le \varepsilon_n + \frac{1}{n} + \frac{A(\gamma)}{n}  \label{eq:particular_x4} 
\end{align}
from \eqref{eq:particular_x} and \eqref{eq:particular_x3}. 

Combining \eqref{eq:mixed_meta_conv2b}, \eqref{eq:e-achievable_rate}, and \eqref{eq:R_n^*} gives 
\begin{align}
R-\gamma \le R_n^* ~~~(\forall n \ge n_1).  \label{eq:e-achievable_rate2}
\end{align}
Then \eqref{eq:particular_x4} implies that there exists a sequence of types $\{ P_{n} \in \mathcal{T}_n \}_{n=n_1}^\infty$ such that
\begin{align}
& \sum_{\ell \in \Omega} w_\ell \vect{1} \left\{  I_{P_n} (X;Y_\ell) \le R   - 3 \gamma  - \frac{1}{n} \log n|\mathcal{T}_n| \right\}  \nonumber \\
&~~~~~~~~~~ \le \varepsilon + \delta_n + \frac{1}{n} + \frac{A(\gamma)}{n}    \label{eq:particular_x6}
\end{align}
holds for all $n \ge n_1$, where the relation $\varepsilon_n \le \varepsilon + \delta_n~(\forall n \ge n_1)$ in \eqref{eq:e-achievable_rate} is used.
Setting $\rho_n := \delta_n + \frac{1}{n} + \frac{A(\gamma)}{n}$, we obtain 
\begin{align}
& \hspace*{-1.5mm} \sum_{\ell \in \Omega} w_\ell \vect{1} \! \left\{  I_{P_n} \! (X;Y_\ell)\le R   {-} {3 \gamma}  {-} \frac{1}{n} \log n|\mathcal{T}_n| \right\}  
\le \varepsilon {+} \rho_n \label{eq:particular_x5}
\end{align}
for $n \ge {n_1}$. 

It can be verified from \eqref{eq:particular_x5} and the definition of $\tilde{R}(\cdot):= \tilde{R}(\cdot|\mathcal{P}(\mathcal{X}))$ that 
\begin{align}
R   - {3 \gamma}  - \frac{1}{n} \log n|\mathcal{T}_n|  \le \tilde{R} (\varepsilon + \rho_n) \label{eq:particular_x5b}
\end{align}
holds for $n \ge {n_1}$.
It is well-known that $|\mathcal{T}_n| \le (n+1)^{|\mathcal{X}|}$ holds by the method of types, and taking the limes superior with respect to $n$ on both sides of \eqref{eq:particular_x5b} yields 
\begin{align}
R -{3 \gamma} &\le \lim_{n\rightarrow \infty} \tilde{R}  (\varepsilon + \rho_n) \, {=}\, \inf_{\delta > 0} \, \tilde{R}  (\varepsilon + \delta). \label{eq:lim_tilde_R}
\end{align} 
The equality in \eqref{eq:lim_tilde_R} is due to Property (d) in Lemma \ref{lem:property_A}.
Since $\gamma > 0$ is an arbitrary constant, \eqref{eq:lim_tilde_R} implies $R \le {\inf_{\delta > 0} \tilde{R} (\varepsilon + \delta)}$, i.e., \eqref{eq:converse_theorem}. 



\subsection{Direct Part}

Direct Part of Theorem \ref{theo:mixed_DMC_e-cap} is stated as follows: 


\begin{e_theo}[Direct Theorem] \label{theo:mixed_DMC_e-cap_direct}
Let $\vect{W}$ be a mixed memoryless channel such that $|\mathcal{X}| < \infty$.
For a fixed $\varepsilon \in [0,1)$, any rate $R$ satisfying
\begin{align}
R <  {\inf_{\delta > 0}} \sup_{ P_X \in \PX } \sup \left\{ R \, \Big| F_w(R|P_X) \le \varepsilon {+ \delta} \right\} \label{eq:achievable_rate}
\end{align}
is $\varepsilon$-achievable. 
\QED
\end{e_theo}

The following lemma is used to prove Direct Part.
\begin{e_lem} \label{lem:limsup_Pr_ell}
Let $P_{X^n}$ be a product distribution of a given $P_X \in \PX$.
Then we have 
\begin{align}
& \limsup_{n\rightarrow \infty}\Pr \left\{ \frac{1}{n} \log \frac{W_\ell^n(Y_\ell^n|X^n)}{(P_XW_\ell)^{\times n}(Y_\ell^n)} \le R + \rho_{\ell,n} \right\}  \nonumber \\
&~~~~~~~~~~~~~~~ \le \vect{1} \left\{ I_{P_X}(X;Y_\ell) \le R + \gamma \right\}~~~(\forall \ell \in \Omega), \label{eq:limsup_Pr_ell}
\end{align}
where $\{\rho_{\ell,n}\ge 0 \} $ denotes an arbitrary sequence such that $\lim_{n \rightarrow \infty} \rho_{\ell,n} =  0$, and $\gamma>0$ denotes an arbitrary constant.
\end{e_lem}
(Proof) ~See Appendix \ref{append:proof_lem_limsup_Pr_ell}.
\QED


We now prove Direct Part. 
Setting 
\begin{align}
R_0:={\inf_{\delta > 0}} \sup_{ P_X \in \PX } \sup \left\{ R \, \Big| F_w(R|P_X) \le \varepsilon {+ \delta} \right\},  \label{eq:R_0}
\end{align}
we shall show that $R:= R_0 - 4\gamma$ is $\varepsilon$-achievable for any $\gamma>0$.

Fix $\gamma > 0$ arbitrarily.
By \eqref{eq:R_0}, we have
\begin{align}
R_0\le \sup_{ P_X \in \PX } \sup \left\{ R \, \Big| F_w(R|P_X) \le \varepsilon + \delta \right\}  \label{eq:R_0b}
\end{align}
for all $\delta> 0$.
For an arbitrarily fixed $\delta>0$, there exists a $P_X^{(\delta)} \in \PX$ such that 
\begin{align}
& \sup_{ P_X \in \PX } \sup \left\{ R \, \Big| F_w(R|P_X) \le \varepsilon + \delta \right\} \nonumber \\
&~~~~~~~~~~
\le \sup \left\{ R \, \Big| F_w \left(R|P_X^{(\delta)} \right) \le \varepsilon + \delta \right\} + \gamma . \label{eq:R_0c}
\end{align}
It follows from \eqref{eq:R_0b} and \eqref{eq:R_0c} that 
\begin{align}
\sup \left\{ R \, \Big| F_w \left(R|P_X^{(\delta)} \right) \le \varepsilon + \delta \right\} \ge R_0 - \gamma > R+2\gamma. \label{eq:R_0d}
\end{align}
Since $F_w(R|P_X^{(\delta)})$ is a non-decreasing function of $R$,  \eqref{eq:R_0d} implies
\begin{align}
 F_w \left(R + 2 \gamma|P_X^{(\delta)} \right) \le \varepsilon + \delta.  \label{eq:R_0e}
\end{align}

On the other hand, by setting $M_n = e^{nR}$, \eqref{eq:error_achievable} holds trivially. 
We now consider the ensemble of random codes for which $n$ symbols of each codeword are randomly chosen according to $P_X^{(\delta)}$ i.i.d. 
That is, 
$P_{X^n}(\vect{x}) = \prod_{i=1}^n P_X^{(\delta)}(x_i)~(\forall \vect{x} \in \mathcal{X}^n)$.
Then Lemma \ref{lem:Feinstein_mixed_channel} guarantees that there exists an $(n,M_n,\varepsilon_n)$ code satisfying 
\begin{align}
& \varepsilon_n \le \sum_{\ell \in \Omega} w_\ell \Pr\left\{ \frac{1}{n} \log \frac{W_\ell^n(Y_\ell^n|X^n)}{(P_X^{(\delta)}W_\ell)^{\times n}(Y_\ell^n)} \le R  \right. \nonumber \\
&~~~~~~~~~~~~~~~~~~~~~\left.  +\, \gamma + \frac{1}{n} \log \frac{1}{w_\ell}\right\} + e^{-n\gamma}. \label{eq:Feinshtein_mixed_channel00}
\end{align}
Taking the limes superior with respect $n$ on both sides in \eqref{eq:Feinshtein_mixed_channel00}, 
\begin{align}
 \limsup_{n \rightarrow \infty} \varepsilon_n \, &\le \sum_{\ell \in \Omega} w_\ell  \limsup_{n \rightarrow \infty} \Pr\left\{ \frac{1}{n} \log \frac{W_\ell^n(Y_\ell^n|X^n)}{(P_X^{(\delta)}W_\ell)^{\times n}(Y_\ell^n)} \right. \nonumber \\
&~~~~~~~~~~\left.\le R + \gamma + \frac{1}{n} \log \frac{1}{w_\ell}\right\} \nonumber\\
 &\le \sum_{\ell \in \Omega} w_\ell \vect{1}\left\{ I_{P_{X}^{(\delta)}}(X;Y_\ell) \le R + 2 \gamma \right\} \label{eq:Feinshtein_mixed_channel2} \\
 &= F_w \left(R+2\gamma|P_{X}^{(\delta)} \right) \le \varepsilon + \delta  \label{eq:epsilon_delta_achievable1}
\end{align}
holds by the sub-additivity of the limes superior.
The inequality in \eqref{eq:Feinshtein_mixed_channel2} is due to Lemma \ref{lem:limsup_Pr_ell}, and the last inequality follows from \eqref{eq:R_0e}. 
Since \eqref{eq:epsilon_delta_achievable1} holds for an arbitrary fixed $\delta>0$,  
\begin{align}
\limsup_{n \rightarrow \infty} \, \varepsilon_n &\le \varepsilon
\end{align}
holds, and thus $R$ is $\varepsilon$-achievable.

\appendices

\section{Proof of Lemma \ref{lem:property_A}}

\subsection{Property (a): Continuity of $A(P_X, \delta)$ in $P_X$}  \label{sec:convexity_A2}

Mutual information $I_{P_X}(X;Y_\ell)$ is uniformly continuous in $P_X$ since the input alphabet $\mathcal{X}$ is finite.
Then we have the following lemma. 
\begin{e_lem} \label{lem:equicontinuity} 
For at most countably many stationary memoryless channels $\{ W_\ell\}_{\ell \in \Omega}$, we have 
\begin{align}
 &\forall \eta {>}0,~\exists \lambda(\eta){>}0,  ~\forall \ell {\in} \Omega,~\forall P_X, P_X' {\in} \mathcal{P}(\mathcal{X}) ~\mbox{s.t.}\nonumber \\
&~ ||P_X - P_X'|| \le \lambda(\eta) ~ \Rightarrow~ |I_{P_X} (X;Y_\ell) - I_{P_X'} (X;Y_\ell)| \le \eta, \label{eq:uniform_continuity}
\end{align}
 where we define 
\begin{align}
||P_X - P_X'|| := \sum_{x \in \mathcal{X}} |P_X(x) - P_X'(x)|, 
\end{align}
the variational distance between $P_X$ and $P_X'$.
\QED
\end{e_lem}
\if01
（Lemma \ref{lem:equicontinuity}の証明）

\eqref{eq:uniform_continuity}が成立しないと仮定する. このとき, 
\begin{align}
& \exists \eta_0 >0, ~\forall \lambda>0,   ~\exists \ell_0 \in \Omega,~\exists P_X, P_X' \in \mathcal{P}(\mathcal{X}) ~~\mbox{s.t.}\nonumber \\
&~ ||P_X - P_X'|| \le \lambda ~~ \mathrm{and}~~ |I_{P_X} (X;Y_{\ell_0}) - I_{{P_X'}} (X;Y_{\ell_0})| > \eta_0 \nonumber 
\end{align}
が成り立つ. 
ここで, $\lambda$は任意なので, $\lambda=\frac{1}{k}$とおく. すなわち, 
\begin{align}
& \exists \eta_0 >0,  ~\exists \ell_0 \in \Omega,~\exists \{(P_k, P_k')\} \in \mathcal{P}(\mathcal{X}) \times  \mathcal{P}(\mathcal{X}) ~~\mbox{s.t.}\nonumber \\
& ||P_k - P_k'|| \le \frac{1}{k} ~~ \mathrm{and}~~ |I_{P_k} (X;Y_{\ell_0}) - I_{{P_k'}} (X;Y_{\ell_0})| > \eta_0 \label{eq:non_uniform_continuity2}
\end{align}
が成り立つ. 
$\{P_k\}$は有界な数列なので, 収束する部分列を持つ. この部分列を$\{P_{\rho(k)}\}$と表わす. 部分例に対しても, 明らかに
$||P_{\rho(k)} - P_{\rho(k)}'|| \le \frac{1}{k} $が成り立っている. 
収束部分列$\{P_{\rho(k)}\}$の極限分布を$P := \lim_{k \rightarrow \infty} P_{\rho(k)}$と表わそう. 
したがって, 任意の$\lambda>0$に対してある$k_0$が存在して, $k \ge k_0$について $|| P_k - P|| \le \lambda$ かつ$ || P_k' - P|| \le \lambda$が成り立つ.  
{一方, $\mathcal{P}(\mathcal{X})$は有界閉集合であり, 相互情報量$I_{P_X}(X;Y_{\ell_0})$は$P_X$の一様連続な関数である. よって
\begin{align}
& \forall \eta >0, ~\exists \lambda_{\ell_0}(\eta)>0,  ~\forall P_X, P_X' \in \mathcal{P}(\mathcal{X})~~\mbox{s.t.}\nonumber \\
& || P_X - P_X'|| \le  \lambda_{\ell_0}(\eta) ~\Rightarrow ~ |I_{P_X} (X;Y_{\ell_0}) - I_{P_X'} (X;Y_{\ell_0})| \le \eta \label{eq:uniform_continuity2}
\end{align}
}このことより, $\forall \eta >0, ~\exists k(\eta)>0,  ~\forall k > k(\eta) ~~\mbox{s.t.}$
\begin{align}
 |I_{P_k} (X;Y_{\ell_0}) - I_{{P}} (X;Y_{\ell_0})| & \le \eta, \nonumber \\
|I_{P_k'} (X;Y_{\ell_0}) - I_{{P}} (X;Y_{\ell_0})| & \le \eta \label{eq:convergence1}
\end{align}
が成り立つことも容易に確かめられる. 

いま, $\eta \le \eta_0/2$なる$\eta$を任意に固定する. \eqref{eq:convergence1}より
\begin{align}
 &|I_{P_k} (X;Y_{\ell_0}) - I_{P_k'} (X;Y_{\ell_0})| \nonumber \\
&\le  |I_{P_k} (X;Y_{\ell_0}) - I_{P} (X;Y_{\ell_0})|  +  |I_{P_k'} (X;Y_{\ell_0}) - I_{P} (X;Y_{\ell_0})|  \nonumber \\
& \le 2\eta \le \eta_0  ~~~~~~(\forall k \ge k(\eta)).
 \end{align}
これは, \eqref{eq:non_uniform_continuity2}に矛盾する. したがって, \eqref{eq:uniform_continuity}が成り立つことが分かる. 

\begin{flushright}
（Lemma \ref{lem:equicontinuity}の証明）
\end{flushright}
\fi




\begin{e_rema}
This lemma holds for an arbitrary family of uniform continuous functions $\{f_\ell(P_X) |\, f_\ell : \mathcal{D} \rightarrow \mathbb{R} \}$, where $\mathcal{D}$ is a compact set in $\mathcal{P}(\mathcal{X})$. A constant $\lambda(\eta)$ in \eqref{eq:uniform_continuity} can be chosen independent of channel index $\ell$ because of the uniform continuity of $f_\ell(P_X)$.
\QED
\end{e_rema}

Fix $\eta > 0$ arbitrarily, and choose  any $P_X, P_X' \in \mathcal{P}(\mathcal{X})$ satisfying $||P_X - P_X'|| \le \lambda(\eta)$. By Lemma \ref{lem:equicontinuity}, we have
\begin{align}
|I_{P_X}(X;Y_\ell)  - I_{P_X'}(X;Y_\ell)| \le \eta~~(\forall \ell \in \Omega). \label{eq:MI}
\end{align}
Since \eqref{eq:MI} implies
\begin{align}
\sum_{\ell \in \Omega} w_\ell \vect{1} \left\{ I_{P_X'}(X;Y_\ell) \le R \right\} \ge \sum_{\ell \in \Omega} w_\ell \vect{1} \left\{ I_{P_X}(X;Y_\ell) \le R {-} \eta \right\}, \nonumber
\end{align}
we have a chain of expansions
\begin{align}
A(P_X', \delta) 
&\le \sup \left\{ R ~\Big| \sum_{\ell \in \Omega} w_\ell \vect{1} \left\{ I_{P_X}(X;Y_\ell) \le R {-} \eta \right\} \le \delta \right\} \nonumber \\ 
&= \sup \left\{ R + \eta ~\big| \sum_{\ell \in \Omega} w_\ell \vect{1} \left\{ I_{P_X}(X;Y_\ell) \le R\right\} \le \delta \right\} \nonumber \\ 
&= A(P_X,\delta) + \eta. \label{eq:ineq4a}
\end{align}
By the same argument, we also have
\begin{align}
A(P_X, \delta) \le A(P_X',\delta) + \eta. \label{eq:ineq4b}
\end{align}
%
%
Since $P_X, P_X'$ are arbitrarily chosen, \eqref{eq:ineq4a} and \eqref{eq:ineq4b} imply  
\begin{align}
|A(P_X, \delta) - A(P_X', \delta)| \le \eta,
\end{align}
 and thus the function $A(P_X, \delta)$ is continuous in $P_X$. 


\subsection{Property (d): Right Continuity of $\tilde{R}(\delta | \mathcal{D})$ in $\delta$} \label{sec:prop_tilde_R}

The function $\tilde{R}(\cdot | \mathcal{D})$ is non-decreasing in $\delta$ because of Property (b) of $A(P_X,\cdot)$.  
Then it is sufficient to show  
\begin{align}
\lim_{k \rightarrow \infty} \tilde{R}(\delta+ \lambda_k | \mathcal{D}) &=  \tilde{R}(\delta | \mathcal{D})\label{eq:right_cont}
\end{align}
by fixing $\delta \in [0,1)$ and a decreasing sequence $\{\lambda_k > 0 | \lambda_1 > \lambda_2 > \cdots  \rightarrow 0\}$ arbitrarily.
We denote by $\mathbb{N}$ the set of all natural numbers.
We assign an index $k \in \mathbb{N}$ to $A(P_X, \delta+ \lambda_k)$ and relabel as $\tilde{A}_k (P_X|\delta) := A(P_X, \delta+ \lambda_k) $.

By the properties of $A(P_X, \delta)$ (Property (a)--(c)), we have the following:
\begin{itemize}
\item[(i)] $\big\{\tilde{A}_k(P_X|\delta)\big\}_{k \in \mathbb{N}} $ is a monotonically decreasing sequence of functions in $k$. 
\item[(ii)] $\lim_{k \rightarrow \infty} \tilde{A}_k (P_X|\delta)  = A(P_X, \delta) $ (pointwise convergence in $P_X$). 
\item[(iii)] $A(P_X, \delta) $ is a continuous function of $P_X$. 
\end{itemize}
Thus, since a monotonically decreasing sequence of functions converges pointwise to a continuous function over a compact set $\mathcal{D}$, Dini's theorem 
holds, and $\big\{\tilde{A}_k (P_X|\delta)\big\}_{k \in \mathbb{N}}$ converge to $A(P_X, \delta)$ uniformly. 
By the uniform convergence, we have
\begin{align}
\lim_{k \rightarrow \infty} \max_{P_X \in \mathcal{D} }\tilde{A}_k (P_X|\delta)  &= \max_{P_X \in \mathcal{D}} \lim_{k \rightarrow \infty} \tilde{A}_k (P_X|\delta) \nonumber \\
&=  \max_{P_X \in \mathcal{D}} A(P_X, \delta) \label{eq:order_change}
\end{align}
(c.f.\ \cite[Lemma 2]{Ahlswede68}). 
By the relation
\begin{align}
\tilde{R}(\delta+\lambda_k | \mathcal{D}) = \max_{P_X \in \mathcal{D}} \tilde{A}_k(P_X|\delta) 
\end{align}
and the definition of $\tilde{R}(\delta|\mathcal{D})$, \eqref{eq:order_change} means \eqref{eq:right_cont}.


\if01
\subsection{式\eqref{eq:order_change}における$\lim$と$\max$の順序交換について}  \label{sec:order_change}

$K$をコンパクトな距離空間とする. $f: K \rightarrow \mathsf{R}$を連続関数として, $f_k: K \rightarrow \mathsf{R}$を連続関数の列とする. 
もし
\begin{align}
\lim_{k \rightarrow \infty} f_k(x) = f(x) 
\end{align}
が一様収束するならば, 
\begin{align}
\max_{x \in K} f(x) = \lim_{k \rightarrow \infty} \max_{x \in K} f_k(x) \label{eq:order_change}
\end{align}
が成り立つことを示す. 以下, 表記の簡単化のため$A:= \max_{x \in K} f(x) $, $B := \lim_{k \rightarrow \infty} \max_{x \in K} f_k(x) $ とおく. 

\noindent
(i) Proof of $A \ge B$ の証明

$x_k^* := \arg \max_{x \in K} f_k(x)$
とおくと, 
\begin{align}
\lim_{k \rightarrow \infty} \max_{x \in K}f_k(x) = \lim_{k \rightarrow \infty} f_k(x_k^*)
\end{align}
の関係が成り立つ. 数列$\{x_k^*\}$は有界なので, 収束する部分列$\{x_{\rho(k)}^*\}$が存在する. 
$c^* = \lim_{k \rightarrow \infty}x_{\rho(k)}^*$をその極限値とする. 当然, $B = \lim_{k \rightarrow \infty} f_{\rho(k)} ( x_{\rho(k)}^*)$が成り立つ. 次のLemma を用いる. 

\begin{e_lem} \label{lem:convergence}
連続関数の列$\{ f_k\}$が連続関数$f$に一様収束するとき, 任意の$c = \lim_{k \rightarrow \infty} x_k$なる収束列$\{x_k\}$ に対して
\begin{align}
\lim_{k \rightarrow \infty} f_k (x_k) = f(c)
\end{align}
が成り立つ. 
\end{e_lem}

Lemma \ref{lem:convergence}より, $B = \lim_{k \rightarrow \infty} f_{\rho(k)} ( x_{\rho(k)}^*) = f(c^*)$
が成り立つ. したがって, 
\begin{align}
B &= \lim_{k \rightarrow \infty} f_{\rho(k)} ( x_{\rho(k)}^*) = f(c^*) \le \max_{x \in K } f(x)
\end{align}
が成り立つ. 

以上の議論より, \eqref{eq:order_change}が成り立ち, $\lim_{k \rightarrow \infty}$と$\max_{x \in K}$の順序が交換できることが分かる.

\noindent
（Lemma \ref{lem:convergence}の証明）

任意の$k \in \mathbb{N}$について
\begin{align}
|f_k(x_k) - f(c)| &\le |f_k (x_k) - f(x_k)| +  |f(x_k) - f(c)| \nonumber \\
 &\le \sup_{x} |f_k(x) - f(x)| + |f(x_k) - f(c)| \label{eq:absolute_diff}
\end{align} 
が成り立つ. $\{f_k\}$が$f$に一様収束することより, 第1項は任意に固定した$\eta >0$に対しある$k_0(\eta)$が存在して, 任意の$k > k_0(\eta)$について$\sup_{x} |f_k(x) - f(x)|  \le \eta$と上から抑えられる. 
一方, 第2項は$c$が$\{ x_k \}$の極限値であることと$f$の連続性より, 任意に固定した$\eta >0$に対しある$\delta(\eta), k_1(\delta(\eta))$が存在して, 任意の$k > k_1(\delta(\eta))$について
\begin{align}
|x_k - c| \le \delta(\eta)    ~~~\mbox{and}~~~ |f(x_k) - f(c)| \le \eta
\end{align}
となる. よって, 任意の$\eta>0$に対して$k(\eta) := \max\{k_0(\eta), k_1(\delta(\eta)) \}$とおくと, \eqref{eq:absolute_diff}により任意の$k > k(\eta)$について
\begin{align}
|f_k(x_k) - f(x)| &\le 2 \eta \label{eq:absolute_diff2}
\end{align} 
となることが分かる. したがって, Lemma が成り立つ. 
\begin{flushright}
（Lemma \ref{lem:convergence}の証明終了）
\end{flushright}

\noindent
(ii) $A \le B$ の証明

$\{f_k\}$が連続関数列で$f$一様収束するので, $f$も連続となる. したがって, $x^* := \arg \max_{x} f(x)$が存在する. $\{f_k\}$が$f$に収束することより, 任意の$\gamma>0$に対して, ある$k_0 > 0$が存在して, 全ての$k \ge k_0$について
\begin{align}
f_k (x^*) + \gamma \ge f(x^*),
\end{align}
が成り立ち, また全ての$k>0$について
\begin{align}
 \max_{x} f_k(x) \ge f_k(x^*)
\end{align}
が成り立つ. これら2式より全ての$k > k_0$について
\begin{align}
\max_{x} f_k(x) \ge f(x^*) - \gamma \label{eq:ineq0}
\end{align}
となる. \eqref{eq:ineq0}の両辺を$\lim_{k \rightarrow \infty}$とると
\begin{align}
B &= \lim_{k \rightarrow \infty} \max_{x} f_k(x) \nonumber \\
&\ge f(x^*) - \gamma =  A -\gamma 
\end{align}
となるが, $\gamma$は任意であったので, 結局$A \le B$が成り立つことが分かる. 

\QED 
\fi

\if01 
\subsection{Diniの定理と一様収束性} \label{sec:Dini_Theorem}

\begin{e_lem}[Dini's Theorem]
$K$をコンパクトな距離空間とする. $f: K \rightarrow \mathsf{R}$を連続関数として, $f_k: K \rightarrow \mathsf{R}$を連続関数の列とする. 
もし
\begin{align}
f_k(x) \le f_{k+1}(x) ~~~~\mbox{for~all}~x \in K \mbox{and~all}~k \in \mathbb{N}
\end{align}
が成り立ち, $\{ f_k\}_{k \in \mathbb{N}}$ が関数$f$に各点収束するならば, 関数列$\{ f_k\}_{k \in \mathbb{N}}$ は関数$f$に一様収束する. 
\end{e_lem}

関数列$\{ f_k\}_{k \in \mathbb{N}}$ が関数$f$に一様収束する必要十分条件は, 数列
\begin{align}
g_k:= \sup_{x \in K} |f_k(x) - f(x)| 
\end{align}
が$\lim_{k \rightarrow \infty} g_k = 0$を満たすことである. 

Diniの定理は, $\varepsilon$-capacity の関数の右連続性を示すのに用いられる. その際, \ref{sec:prop_tilde_R}節における記号と本節の記号は以下のように対応する. 
\begin{itemize}
\item $K \leftrightarrow \mathcal{P}(\mathcal{X})$ ($\mathcal{P}(\mathcal{X})$はコンパクト集合)
\item $f_k \leftrightarrow \tilde{A}_k(\cdot|\varepsilon)$
\item $f \leftrightarrow A(\cdot,\varepsilon)$
\end{itemize}

\fi


\section{Proof of Lemma \ref{lem:alternative_express}} \label{sec:proof_alternative_express}

Fix an input probability distribution $P_X \in \PX$ arbitrarily.
It is easily verified that the l.h.s.\ of \eqref{eq:alternative_express} can be expressed as 
\begin{align}
&\hspace*{-1mm} \sup\left\{ R | \, F_w(R|P_X) \le \varepsilon \right\} \nonumber \\
&~~= \sup \left\{ R \, \Big| \, \sum_{\ell} w_\ell \vect{1} \{ I_{P_X} (X;Y_\ell) < R\} \le \varepsilon \right\} \nonumber \\
&~~= \sup \left\{ R \, \Big| \, \sum_{\ell} w_\ell \vect{1} \{ I_{P_X} (X;Y_\ell) \ge R\} \ge 1- \varepsilon \right\}. \label{eq:alternative_A}
\end{align}
Therefore, defining 
\begin{align}
A(\varepsilon|P_X) &:= \sup \left\{ R \, \Big| \, \sum_{\ell} w_\ell \vect{1} \{ I_{P_X} (X;Y_\ell) \ge R\} \ge 1- \varepsilon \right\}, \label{eq:set_A} \\
B(\varepsilon|P_X) &:= \sup_{\{S \subseteq \Omega | \, w(S) \ge 1 -\varepsilon\}} \inf_{\ell \in S} I_{P_X}(X;Y_\ell),  \label{eq:set_B} 
\end{align}
we shall show $A(\varepsilon|P_X)  = B(\varepsilon|P_X)$. 

\noindent
(i) Proof of $A(\varepsilon|P_X)  \ge B(\varepsilon|P_X)$: 

Set $R_0 := B(\varepsilon|P_X)$. 
By the definition of $B(\varepsilon|P_X)$, for any fixed $\gamma > 0$, there exists $S_0 \subseteq \Omega$ satisfying $w(S_0) \ge 1- \varepsilon$ and
\begin{align}
R_0 &\le \inf_{k \in S_0} I_{P_X}(X;Y_k) +\gamma. \label{eq:ineq_R0}
\end{align}
Also, by the definition of infimum, we have a chain of inequalities
\begin{align}
&\inf_{k \in S_0} I_{P_X}(X;Y_k) \nonumber \\
&~~= \sup \left\{ R \, \Big| \, I_{P_X} (X;Y_\ell) \ge R ~(\forall \ell \in S_0) \right\} \nonumber  \\
&~~= \sup  \left\{ R \, \Big| \, I_{P_X} (X;Y_\ell) \ge R ~~(\forall \ell \in S_0), \right. \nonumber \\
&~~~~~~~~~~~~~\left.~~ \sum_{\ell \in S_0} w_\ell \vect{1} \{ I_{P_X} (X;Y_\ell) \ge R\} \ge 1- \varepsilon \right\} \nonumber  \\
&~~\le \sup  \left\{ R \, \Big| \,  \sum_{\ell \in S_0} w_\ell \vect{1} \{ I_{P_X} (X;Y_\ell) \ge R\} \ge 1- \varepsilon \right\} \nonumber  \\
&~~= A (\varepsilon|P_X). \label{eq:ineq_R0_2}
\end{align}
By \eqref{eq:ineq_R0} and \eqref{eq:ineq_R0_2}, we have
\begin{align}
R_0 -\gamma &\le A(\varepsilon|P_X), \label{eq:ineq_R0_3}
\end{align}
concluding $R_0 \le A(\varepsilon|P_X)$ since $\gamma>0$ is fixed arbitrarily,   

\noindent
(ii) Proof of $A(\varepsilon|P_X)  \le B(\varepsilon|P_X)$:


We define the set 
\begin{align}
S(\rho) := \left\{ \ell \in \Omega \big| I_{P_X}(X;Y_{\ell}) \ge \rho \right\} 
\end{align}
for $\rho> 0$. It should be noticed that 
\begin{align}
w \left(S(\rho_1) \right) \ge w \left(S(\rho_2) \right) 
\end{align}
for any $0< \rho_1 \le \rho_2$. 

Consider the value $\rho^* > 0$ satisfying the following conditions: 
\begin{align}
w \left(S(\rho^*-\gamma) \right) &\ge 1 -\varepsilon ~~~(\forall \gamma > 0), \\
w \left(S(\rho) \right) &< 1 -\varepsilon ~~~(\forall \rho > \rho^*). \label{eq:cond_rho^*}
\end{align}
For an arbitrarily fixed $\eta >0$, we have $S(\rho^*+\eta) \subset S(\rho^*-\eta) $ and
\begin{align}
\sum_{\ell \in \Omega} w_\ell \vect{1} \left\{ I_{P_X} (X;Y_\ell) \ge \rho^* + \eta \right\} = w(S(\rho^*+\eta) ) < 1 - \varepsilon \label{eq:ineq1}
\end{align}
from \eqref{eq:cond_rho^*}.  
Since every $R > 0$ such that 
\begin{align}
\sum_{\ell \in \Omega} w_\ell \vect{1} \left\{ I_{P_X} (X;Y_\ell) \ge R \right\}  < 1 - \varepsilon
\end{align}
 satisfies $R \ge A(\varepsilon|P_X)$ by the definition of $A(\varepsilon|P_X)$, \eqref{eq:ineq1} implies
\begin{align}
A(\varepsilon|P_X) \le  \rho^* + \eta. \label{eq:ineq_A}
\end{align}
Meanwhile, we have 
\begin{align}
\rho^* - \eta \le \inf_{k \in S(\rho^*-\eta)} I_{P_X} (X;Y_k) \le B(\varepsilon|P_X), \label{eq:ineq_B}
\end{align}
where the first inequality follows from the definition of $S(\rho)$, and the second one follows from the fact $w(S(\rho^*-\eta)) \ge 1 -\varepsilon$ and the definition of $B(\varepsilon|P_X)$.
It follows from  \eqref{eq:ineq_A} and \eqref{eq:ineq_B} that 
\begin{align}
A(\varepsilon|P_X) \le  B(\varepsilon|P_X) + 2\eta \label{eq:ineq_C}
\end{align}
holds. Since $\eta> 0$ is arbitrarily fixed, it concludes
$A(\varepsilon|P_X) \le B(\varepsilon|P_X)$.

\section{Proof of Lemma \ref{lem:mixed_meta_conv}} \label{sec:lem_proof2}

{Suppose that the decoder $\xi: \mathcal{Y}^n \rightarrow \{1,  \ldots, M_n\}$ attains the error probability $\varepsilon_n$ without loss of generality.}
Setting $\xi_\ell  = \xi~(\forall \ell \in \Omega)$, and denoting by $ \xi_\ell^{\rm ML}$ the maximum likelihood decoder over $W_\ell^n$, we have 
\begin{align}
\varepsilon_n &=  1 - \frac{1}{M_n} \sum_{i=1}^{M_n} \sum_{\vect{y} \in \mathcal{Y}^n} W^n(\vect{y} | \phi(i)) \xi(i| \vect{y}) \nonumber \\
 & = \sum_{\ell \in \Omega} w_\ell  \left\{ 1 - \frac{1}{M_n} \sum_{i=1}^{M_n}  \sum_{\vect{y} \in \mathcal{Y}^n} W_\ell^n(\vect{y} | \phi(i)) \xi_\ell (i| \vect{y}) \right\} \label{eq:mixed_LB} \\
  & \ge  \sum_{\ell \in \Omega} w_\ell  \left\{ 1 - \frac{1}{M_n} \sum_{i=1}^{M_n}  \sum_{\vect{y} \in \mathcal{Y}^n} W_\ell^n(\vect{y} | \phi(i)) \xi_\ell^{\rm ML} (i| \vect{y}) \right\}. \label{eq:mixed_LB2} 
\end{align} 
Here, the terms inside the brace $\{\cdot\}$ in \eqref{eq:mixed_LB} corresponds to the average error probability $\varepsilon_n^{(\ell)}$ of the decoder $\xi_\ell = \xi$ over $W_\ell^n$, and the terms inside the brace $\{ \cdot\}$ in \eqref{eq:mixed_LB2} denotes the average error probability $\varepsilon_\ell^{\rm ML}$ of the maximum likelihood decoder
 $\xi_\ell^{\rm ML}$.
The inequality in \eqref{eq:mixed_LB2} follows from the fact that the maximum likelihood decoder attains the minimum error probability among all decoders over $W_\ell^n$. 
The probability $\varepsilon_\ell^{\rm ML}$  can be evaluated by using $\alpha_\beta(\cdot,\cdot)$ according to the following lemma shown by Vazquez-Vilar et al.\ \cite{VTGM2013}.

\begin{e_lem}[Vazquez-Vilar et al.\ \cite{VTGM2013}] \label{lem:tightness_meta_conv}
For a given code $\Code$ of length $n$ and the number of codewords $M_n$, the average error probability of the maximum likelihood decoder over the channel $W_\ell^n$ is given by
\begin{align}
\varepsilon_\ell^{\rm ML} =  \sup_{Q_{Y_\ell^n}} \, \alpha_{\frac{1}{M_n}} (P_{X^n} W_\ell^n, P_{X^n} Q_{Y_\ell^n}). \label{eq:meta_conv_tightness}
\end{align}
Here, $P_{X^n}$ denotes the uniform distribution on $\Code$, and the max on the r.h.s.\ is taken over all probability measures on $\mathcal{Y}^n$. 
\QED
\end{e_lem}

Applying Lemma \ref{lem:tightness_meta_conv} for \eqref{eq:mixed_LB2} yields 
\begin{align}
\varepsilon_n &\ge \sum_{\ell \in \Omega} w_\ell \, \sup_{Q_{Y_\ell^n}} \alpha_{\frac{1}{M_n}} (P_{X^n} W_\ell^n, P_{X^n} Q_{Y_\ell^n}). \label{eq:mixed_LB3} 
\end{align}
Thus,  \eqref{eq:mixed_meta_conv_alpha} holds.

%
By using a duality of $(\alpha, \beta_{\alpha})$ and $(\alpha_\beta, \beta)$ in simple hypothesis testing, \eqref{eq:meta_conv_tightness} implies
\begin{align}
\frac{1}{M_n} &\ge \beta_{\varepsilon_\ell^{\rm ML}} (P_{X^n} W_\ell^n, P_{X^n} Q_{Y_\ell^n}) 
\label{eq:alpha_ineq4}
\end{align}
for every fixed $Q_{Y_\ell^n}$,
which can be easily verified by considering the region of possible pairs of $(\alpha,\beta)$ (c.f.\ \cite[Figure 3.1]{Lehmann-Romano2005}). 
Since $\varepsilon_\ell^{\rm ML} \le \varepsilon_n^{(\ell)}$, we have 
\begin{align}
\beta_{\varepsilon_\ell^{\rm ML}} (P_{X^n} W_\ell^n, P_{X^n} Q_{Y_\ell^n}) \ge \beta_{\varepsilon_n^{(\ell)}} (P_{X^n} W_\ell^n, P_{X^n} Q_{Y_\ell^n}) 
\end{align}
for any given $Q_{Y_\ell^n}$, yielding the inequality
\begin{align}
\frac{1}{M_n} & \ge  \sup_{Q_{Y_\ell^n}} \beta_{\varepsilon_n^{(\ell)}} (P_{X^n} W_\ell^n, P_{X^n} Q_{Y_\ell^n}), \label{eq:alpha_ineq5}
\end{align}
from \eqref{eq:alpha_ineq4}.
Lower bounding the r.h.s. of \eqref{eq:alpha_ineq5} by fixing some $\{Q_{Y_\ell^n}\}_{\ell \in \Omega}$ and taking the mixture with the mixing ratio $\{w_\ell\}_{\ell \in \Omega}$ conclude that \eqref{eq:mixed_meta_conv_beta} holds.

\section{Proof of Lemma \ref{lem:relaxed_meta_conv}} \label{sec:lem_proof3}

We first set $R_0 := - \log \sum_{\ell \in \Omega} w_\ell \beta_{\varepsilon_\ell}(P_\ell, Q_\ell) + \log \delta $ and denote by $\xi_\ell^*$ a probabilistic test that attains $\beta_{\varepsilon_\ell}(P_\ell, Q_\ell)$ in the hypothesis testing \eqref{eq:HT_ell}.
We denote by $T^* \in \{H_0,H_1 \}$ the random variable corresponding to the hypothesis estimated by this test.
That is, $P_\ell\{T^* = H_1 \}=\varepsilon_\ell$ and  $Q_\ell  \left\{ T^* = H_0\right\}  = \beta_{\varepsilon_\ell}(P_\ell,Q_\ell)$ hold by the well-known Neyman-Pearson lemma.
Then a standard bounding technique gives
\begin{align}
1-\varepsilon_\ell &= P_\ell \{ T^* = H_0 \} \nonumber\\
 &=  P_\ell \left\{ T^* = H_0, \log \frac{P_\ell(Z_\ell)}{Q_\ell(Z_\ell)} > R_0 \right\} \nonumber \\
&~~~~~ +   P_\ell \left\{ T^* = H_0, \log \frac{P_\ell(Z_\ell)}{Q_\ell(Z_\ell)}\le R_0 \right\}  \nonumber\\
  &\le  P_\ell \left\{ \log \frac{P_\ell(Z_\ell)}{Q_\ell(Z_\ell)} > R_0 \right\} +  e^{R_0} Q_\ell \left\{ T^* = H_0\right\},  
\end{align}
and this implies 
\begin{align}
\varepsilon_\ell &\ge  P_\ell \left\{ \log \frac{P_\ell(Z_\ell)}{Q_\ell(Z_\ell)} \le R_0 \right\} -  e^{R_0} \beta_{\varepsilon_\ell} (P_\ell, Q_\ell) ~~(\forall \ell \in \Omega)
\end{align}
by the definition of $\xi_\ell^*$.  
Since $\{ \varepsilon_\ell \}_{\ell \in \Omega}$ satisfies $\sum_{\ell \in \Omega} w_\ell \varepsilon_\ell = \varepsilon$, taking the mixture of both sides with $\{ w_\ell \}_{\ell \in \Omega}$ yields
\begin{align}
\varepsilon &\ge \sum_{\ell \in \Omega} w_\ell  P_\ell \left\{ \log \frac{P_\ell(Z_\ell)}{Q_\ell(Z_\ell)} \le R_0 \right\} -  e^{R_0}  \sum_{\ell \in \Omega} w_\ell  \beta_{\varepsilon_\ell} (P_\ell, Q_\ell) \nonumber\\ 
&= \sum_{\ell \in \Omega} w_\ell  P_\ell \left\{ \log \frac{P_\ell(Z_\ell)}{Q_\ell(Z_\ell)} \le R_0 \right\} - \delta. \label{eq:meta_conv4}
\end{align}
Here, the equality simply follows from the definition of $R_0$. 
\eqref{eq:meta_conv4} indicates
\begin{align}
R_0 &\le \sup \left\{ R \, \Big| \, \sum_{\ell \in \Omega} w_\ell  P_\ell \left\{ \log \frac{P_\ell(Z_\ell)}{Q_\ell(Z_\ell)} \le R \right\} \le \varepsilon + \delta \right\} \nonumber \\
&=D_{\rm s}^{\varepsilon+ \delta} (\{P_\ell \} || \{Q_\ell \}), \label{eq:meta_conv5}
\end{align}
and thus \eqref{eq:mixed_meta_conv2} holds. 

\section{Proof of Lemma \ref{lem:limsup_Pr_ell}} \label{append:proof_lem_limsup_Pr_ell}

Fix $\gamma>0$ and $\{\rho_{\ell,n}\ge 0 \} $ arbitrarily. 
We define 
\begin{align}
&B_\ell^{(n)}(\gamma) := \left\{ (\vect{x},\vect{y}) \in \mathcal{X}^n \times \mathcal{Y}^n \Big|  \right. \nonumber \\
&~~~~~~~~~\left. \left| \frac{1}{n} \log \frac{W_\ell^n(\vect{y}|\vect{x})}{(P_XW_\ell)^{\times n}(\vect{y})} - I_{P_X}(X;Y_\ell) \right| \le \gamma \right\}
\end{align}
and use a standard bounding technique for each $\ell \in \Omega$ to expand 
\begin{align}
&\hspace*{-2mm} \Pr \left\{ \frac{1}{n} \log \frac{W_\ell^n(Y_\ell^n|X^n)}{(P_XW_\ell)^{\times n}(Y_\ell^n)} \le R + \rho_{\ell,n} \right\} \nonumber \\
&\hspace*{-4mm} ~~~\le \Pr\left\{ \frac{1}{n} \log \frac{W_\ell^n(Y_\ell^n|X^n)}{(P_XW_\ell)^{\times n}(Y_\ell^n)} \le R + \rho_{\ell,n} , \right. \nonumber \\
&\hspace*{-4mm} ~~~~~~\left. (X^n,Y_\ell^n) \! \in \!  B_\ell^{(n)}(\gamma)\right\} \! + \!  \Pr\left\{  (X^n,Y_\ell^n) \! \not\in \! B_\ell^{(n)}(\gamma)\right\}. \label{eq:Pr_ell_UB1}
\end{align}

The random variable $\log \frac{W_\ell^n(Y_\ell^n|X^n)}{(P_XW_\ell)^{\times n}(Y_\ell^n)} $ is a sum of independent random variables.
Then, similar to Lemma \ref{lem:Chebyshev_ineq}, we can apply the Chebyshev inequality to the second term of \eqref{eq:Pr_ell_UB1}  and obtain
\begin{align}
\Pr\left\{  (X^n,Y_\ell^n) \not\in B_\ell^{(n)}(\gamma)\right\} \le \frac{A(\gamma)}{n} \label{eq:Pr_ell_UB1_2nd}
\end{align}
with some constant $A(\gamma) \ge  0$. 
It should be noticed that the variance of the random variable $\frac{1}{n}\log \frac{W_\ell^n(Y_\ell^n|X^n)}{(P_XW_\ell)^{\times n}(Y_\ell^n)}$ is uniformly bounded in $\ell$ because $\mathcal{X}$ is finite (c.f.\ \cite[Remark 3.1.1]{Han2003}), and thus a constant $A(\gamma)$ can be chosen independently of $\ell$.
On the other hand, the first term of \eqref{eq:Pr_ell_UB1} can be bounded as
\begin{align}
& \Pr\left\{ \frac{1}{n} \log \! \frac{W_\ell^n(Y_\ell^n|X^n)}{(P_XW_\ell)^{\times n}(Y_\ell^n)} \le R + \rho_{\ell,n} , (X^n,Y_\ell^n) {\in} B_\ell^{(n)}\!(\gamma)\right\} \nonumber \\
&~~~~~  \le  \vect{1} \left\{ I_{P_X}(X;Y_\ell) -\gamma \le R + \rho_{\ell,n}  \right\}, \label{eq:Pr_ell_UB1_1st}
\end{align}
 which can be verified as follows:
(i) If $I_{P_X}(X;Y_\ell) -\gamma \le R + \rho_{\ell,n}$, \eqref{eq:Pr_ell_UB1_1st} holds trivially because $\vect{1}\{I_{P_X}(X;Y_\ell) -\gamma \le R + \rho_{\ell,n}\} = 1$, and 
(ii)  If $I_{P_X}(X;Y_\ell) -\gamma > R + \rho_{\ell,n}$, 
we have 
\begin{align}
& \Pr\left\{ \frac{1}{n} \log \frac{W_\ell^n(Y_\ell^n|X^n)}{(P_XW_\ell)^{\times n}(Y_\ell^n)} \le R + \rho_{\ell,n} , \right. \nonumber \\
&~~~~~~~~~~~~\left.(X^n,Y_\ell^n)  \in B_\ell^{(n)}(\gamma)\right\} =0
\end{align}
because   
\begin{align}
I(X;Y_\ell) -\gamma \le \frac{1}{n} \log \frac{W_\ell^n(\vect{y}|\vect{x})}{(P_XW_\ell)^{\times n}(\vect{y})}  \nonumber 
\end{align}
for all $(\vect{x},\vect{y})  \in B_\ell^{(n)}(\gamma) $.
This implies that \eqref{eq:Pr_ell_UB1_1st} also holds. 

By \eqref{eq:Pr_ell_UB1}--\eqref{eq:Pr_ell_UB1_1st}, we obtain
\begin{align}
& \Pr \left\{ \frac{1}{n} \log \frac{W_\ell^n(Y_\ell^n|X^n)}{(P_XW_\ell)^{\times n}(Y_\ell^n)} \le R + \rho_{\ell,n} \right\} \nonumber \\
&~~~~~~~~\le \vect{1} \left\{ I_{P_X}(X;Y_\ell)  \le R + \rho_{\ell,n} + \gamma \right\} + \frac{A(\gamma)}{n}. 
 \label{eq:Pr_ell_UB2}
\end{align}
Taking the limes superior with respect $n$ on both sides yields
\begin{align}
& \limsup_{n \rightarrow \infty} \Pr \left\{ \frac{1}{n} \log \frac{W_\ell^n(Y_\ell^n|X^n)}{(P_XW_\ell)^{\times n}(Y_\ell^n)} \le R + \rho_{\ell,n} \right\} \nonumber \\
& ~~~~~~~\le \vect{1} \left\{ I_{P_X}(X;Y_\ell)  \le R + 2 \gamma \right\}, 
 \label{eq:Pr_ell_UB5}
\end{align}
concluding \eqref{eq:limsup_Pr_ell} since $\gamma>0$ is arbitrary.  

\if01
\section{コスト制約付き$\varepsilon$-符号化}

同一タイプアンサンブルを用いるときのDirect Theorem の証明

{
\begin{e_lem} \label{lem:limsup_Pr_ell3}
任意に固定した数列$\{\rho_{\ell,n} \ge 0| \lim_{n \rightarrow \infty} \rho_{\ell,n} =  0\}$と固定した$P_X \in \PX$に対して, $||P_n^o - P_X|| \le \frac{|\mathcal{X}|}{n}$を満たすタイプの列$\{ P_n^o \in \mathcal{T}_n\}_{n=1}^\infty$が与えられたとする. このとき, タイプ$P_n^o$を持つ任意の系列$\vect{x}$と任意の$\gamma > 0$について
\begin{align}
&\limsup_{n\rightarrow \infty}\Pr \left\{ \frac{1}{n} \log \frac{W_\ell^n(Y_\ell^n|X^n)}{(P_n^oW_\ell)^{\times n}(Y_\ell^n)} \le R + \rho_{\ell,n} \Big| X^n = \vect{x} \right\} \nonumber \\
&~~~\le \vect{1} \left\{ I_{P_X}(X;Y_\ell) \le R + \gamma \right\}~~~(\forall \ell=1,2,\cdots) \label{eq:limsup_Pr_ell2}
\end{align}
が成り立つ. 
\end{e_lem}
(Proof)~~証明は\ref{sec:lem_proof4}節で与える. 
\QED
}

一方, $M_n = e^{nR}$とおくと, \eqref{eq:error_achievable}が自明に成立する. 
長さ$n$を固定したとき, $||P_n^o - P_X^{(\delta)}|| \le \frac{|\mathcal{X}|}{n}$を満たすタイプ$P_n^o \in \mathcal{T}_n$が必ず存在する. 
このとき, タイプ$P_n^o $を持つ系列集合から一様ランダムに$M_n$個の符号語を選択するランダム符号化を考えると, Lemma \ref{lem:Feinstein_mixed_channel}より次式を満たす$(n,M_n,\varepsilon_n)$符号が存在する. 
\begin{align}
& \varepsilon_n \le \sum_{\ell \in \Omega} w_\ell \Pr\left\{ \frac{1}{n} \log \frac{W_\ell^n(Y_\ell^n|X^n)}{(P_n^oW_\ell)^{\times n}(Y_\ell^n)} \le \frac{1}{n} \log M_n + \gamma \right. \nonumber \\
&~~~~\left.+ \frac{1}{n} \log|\mathcal{T}_n| + \frac{1}{n} \log \frac{1}{w_\ell}\right\} + e^{-n\gamma}.
\end{align}
したがって, 
\begin{align}
& \varepsilon_n \le \sum_{\ell \in \Omega} w_\ell \Pr\left\{ \frac{1}{n} \log \frac{W_\ell^n(Y_\ell^n|X^n)}{(P_n^oW_\ell)^{\times n}(Y_\ell^n)} \le R + \gamma \right. \nonumber \\
&~~~~\left.+ \frac{1}{n} \log|\mathcal{T}_n| + \frac{1}{n} \log \frac{1}{w_\ell}\right\} + e^{-n\gamma}. \label{eq:Feinshtein_mixed_channel3}
\end{align}
\eqref{eq:Feinshtein_mixed_channel2}の両辺$n$に関する上極限をとると, 上極限の劣加法性より
\begin{align}
 \limsup_{n \rightarrow \infty} \varepsilon_n \, &\le \sum_{\ell \in \Omega} w_\ell  \limsup_{n \rightarrow \infty} \Pr\left\{ \frac{1}{n} \log \frac{W_\ell^n(Y_\ell^n|X^n)}{(P_n^oW_\ell)^{\times n}(Y_\ell^n)} \right. \nonumber \\
&~~~~\left. \le R + \gamma + \frac{1}{n} \log|\mathcal{T}_n| + \frac{1}{n} \log \frac{1}{w_\ell}\right\} \nonumber\\
 &\le \sum_{\ell \in \Omega} w_\ell \vect{1}\left\{ I_{P_{X}^{(\delta)}}(X;Y_\ell) \le R + 2 \gamma \right\} \label{eq:Feinshtein_mixed_channel4} \\
 &= F_w \left(R+2\gamma|P_{X}^{(\delta)} \right) \le \varepsilon + \delta ~~~~~(\forall \delta>0) \label{eq:epsilon_delta_achievable2}
\end{align}
が成り立つ. \eqref{eq:Feinshtein_mixed_channel2}の不等号はLemma \ref{lem:limsup_Pr_ell3}により, 最後の不等式は\eqref{eq:R_0e}による. 
任意の$\delta>0$に対して\eqref{eq:epsilon_delta_achievable2}が成り立つので, 
\begin{align}
\limsup_{n \rightarrow \infty} \, \varepsilon_n &\le \varepsilon.
\end{align}
何故ならば, 仮に$\limsup_{n \rightarrow \infty} \varepsilon_n > \varepsilon$が成り立つと仮定すると, 十分小さい$\delta_0 > 0$をとれば$\limsup_{n \rightarrow \infty} \varepsilon_n > \varepsilon + \delta_0$となるが, このことは\eqref{eq:epsilon_delta_achievable2}に矛盾するためである. 
したがって, $R$は$\varepsilon$-achievableであることが示される.


\fi

\if01
\subsection{Lemma \ref{lem:relaxed_meta_conv}}

仮説検定
\begin{align}
H_0: Z \sim P~~\mbox{vs.}~~~ H_1: Z \sim Q
\end{align}
に対して, Neyman-Pearson 検定を$\xi^*$とする. すなわち, 
\begin{align}
\alpha &= \E_{P} \{ \xi^*(Z) \}, \nonumber \\
\beta_{\alpha}(P, Q) &= \E_{Q} \{ \xi^*(Z) \}\nonumber  
\end{align}
が成り立つものとする. 

\subsection{Lemma }

Converse Theorem において, 出力分布$Q_{Y^n}$を以下のように構成する. いま, $W_\ell^n \ll Q_{Y^n}$ なる任意の出力分布$Q_{Y_\ell^n}(\vect{y})~(\ell = 1,2,\cdots,L)$に対し, 
\begin{align}
Q_{Y^n}(\vect{y}) = \sum_{\ell \in \Omega} w_\ell Q_{Y_\ell^n}(\vect{y}), ~~\vect{y} \in \mathcal{Y}^n \label{eq:Q_Y}
\end{align}
とおく. 混合通信路$W^n$は得られる出力分布$Q_{Y^n}$に対して絶対連続になる.

\begin{e_lem} \label{lem:improved_mixed_LB}
任意の$(n,M_n,\varepsilon_n)$符号は以下を満たす. 
\begin{align}
\varepsilon_n \ge \sum_{\ell \in \Omega} w_\ell P_{X^nY_\ell^n} \left[ \frac{W_\ell^n(Y_\ell^n|X^n)}{Q_{Y_\ell^n}(Y_\ell^n)} \le \gamma \right] - \gamma \beta_{1-\varepsilon}(P_{X^n}W_{Y^n}, P_{X^n}Q_{Y^n}). \label{eq:improved_mixed_LB}
\end{align}
ここで, $Q_{Y_\ell^n}$は任意の出力分布であり, 
\begin{align}
Q_{Y^n} (\vect{y}) = \sum_{\ell \in \Omega} w_\ell Q_{Y_\ell^n}(\vect{y})~~(\forall \vect{y} \in \mathcal{Y}^n)
\end{align}
とする. 
\end{e_lem}

$\mathcal{Z}$上の2つの確率分布$P, Q$の間の2元仮説検定問題を考える. 判定器は$Z$を観測した元で, 確率的判定器$P_{T|Z}$により$T \in \{0, 1\}$ を出力する. $T=1$のとき$P$が真と判定され, $T=0$のとき$P$が真と判定される. 第1種の誤り率を$1-\alpha$以下に抑えた元で, 最適な第2種誤り率を
\begin{align}
\beta_{\alpha}(P,Q) = \min_{P_{T|Z} : ~ P(T=1) \ge \alpha} Q(T=1)
\end{align}
と定義する. ただし, 
\begin{align}
P(T=1) &:= \sum_{z\in \mathcal{Z}} P_{T|Z}(1|z) P(z), \label{eq:P_T=1} \\
Q(T=1) &:= \sum_{z\in \mathcal{Z}} P_{T|Z}(1|z) Q(z)  \label{eq:Q_T=1}
\end{align}
とする. 
Polyanskiyらは任意の$\gamma \in \mathsf{R}$について, 以下の不等式を示している. 
\begin{align}
\alpha \le P\left[ \frac{P(Z)}{Q(Z)} >  \gamma \right] + \gamma \beta_{\alpha}(P,Q).
\end{align}

これに対し, 混合情報源$P, Q$に対して, 以下のLemma が成り立つ. 
\begin{e_lem}
可算無限個の情報源から成る混合情報源$P, Q$間の仮説検定問題を考える. このとき, 任意の$\gamma \in \mathsf{R}$について
\begin{align}
\alpha \le \sum_{\ell=1} w_\ell P_\ell \left[ \frac{P_\ell(Z_\ell)}{Q_\ell(Z_\ell)} > \gamma \right] + \gamma \beta_{\alpha}(P,Q). \label{eq:mix_alpha_ineq}
\end{align}
が成り立つ. 
\end{e_lem}
(Proof)~~$P_{T|Z}$を$\beta_\alpha(P,Q)$を与える判定器とする. このとき, $Q(T=1) = \beta_\alpha(P,Q)$が成り立っている. この判定器を用いて, 
\begin{align}
\alpha &\le \sum_{\ell=1} w_\ell P_\ell (T=1) \nonumber \\
&= \sum_{\ell=1} w_\ell \left\{ P_\ell\left[ T=1, \frac{P_\ell(Z_\ell)}{Q_\ell(Z_\ell)} > \gamma \right] + P_\ell\left[ T=1, \frac{P_\ell(Z_\ell)}{Q_\ell(Z_\ell)} \le \gamma \right]  \right\} \nonumber \\
&\le \sum_{\ell=1} w_\ell \left\{ P_\ell\left[ \frac{P_\ell(Z_\ell)}{Q_\ell(Z_\ell)} > \gamma \right] + P_\ell\left[ T=1, \frac{P_\ell(Z_\ell)}{Q_\ell(Z_\ell)} \le \gamma \right]  \right\}  \label{eq:mix_alpha_ineq1}
\end{align}
{\textbf{（ここまでは, $\gamma_\ell$として良い. しかし, このおき方が有効かは分からない. ）}}\eqref{eq:mix_alpha_ineq1}のthe r.h.s.\ 第2項について, 
\begin{align}
P_\ell\left[ T=1, \frac{P_\ell(Z_\ell)}{Q_\ell(Z_\ell)} \le \gamma \right]  &= \sum_{z} P_{T|Z}(1|z) P_\ell(z) \vect{1} \left[P_\ell(z)\le \gamma Q_\ell(zl) \right] \nonumber \\
&\le \gamma  \sum_{z} P_{T|Z}(1|z) Q_\ell(z) = \gamma Q_\ell(T=1) \label{eq:mix_alpha_ineq_2nd}
\end{align}
が成り立つ. ただし, 
\begin{align}
 Q_\ell(T=1):= \sum_{z\in \mathcal{Z}} P_{T|Z}(1|z) Q_\ell(z)
\end{align}
とおいている. このとき, \eqref{eq:Q_T=1}より
\begin{align}
Q(T=1) = \sum_{\ell=1} w_\ell Q_\ell(T=1)
\end{align}
が成り立っていることに注意しよう. \eqref{eq:mix_alpha_ineq_2nd}を\eqref{eq:mix_alpha_ineq1}に代入して, 
\begin{align}
\alpha &\le \sum_{\ell=1} w_\ell \left\{ P_\ell\left[ \frac{P_\ell(Z_\ell)}{Q_\ell(Z_\ell)} > \gamma \right] + \gamma Q_\ell(T=1) \right\} \nonumber \\
& = \sum_{\ell=1} w_\ell P_\ell\left[ \frac{P_\ell(Z_\ell)}{Q_\ell(Z_\ell)} > \gamma \right] + \gamma \sum_{\ell=1} w_\ell Q_\ell(T=1) \nonumber \\
&=  \sum_{\ell=1} w_\ell P_\ell\left[ \frac{P_\ell(Z_\ell)}{Q_\ell(Z_\ell)} > \gamma \right] + \gamma Q(T=1) 
\end{align}
を得る. ここで, $Q(T=1) = \beta_\alpha(P,Q)$が成り立っていることから, 結局\eqref{eq:mix_alpha_ineq}が得られる. 
\QED

\eqref{eq:meta_LB}に注意すると, 以下の定理が示される. 
\begin{e_theo} \label{theo:improved_mixed_LB}
任意の$(n,M_n,\varepsilon_n)$符号は以下を満たす. 
\begin{align}
\varepsilon_n \ge \sum_{\ell=1} w_\ell P_{X^n Y_\ell^n} \left[ \frac{W_\ell^n(Y_\ell^n | X^n)}{Q_{Y_\ell^n}(Y_\ell^n)} \le \gamma \right] - \gamma/M_n. \label{eq:mix_alpha_ineq4}
\end{align}
ただし, $Q_{Y_\ell^n}$は任意の出力分布とする. 
\QED
\end{e_theo}

\eqref{eq:mixed_LB1}を用いたとき, 得られる下界式は\eqref{eq:HN_bound}より
\begin{align}
\varepsilon_n &\ge \sum_{\ell \in \Omega} w_\ell P_{X^nY_\ell^n} \left\{ \frac{1}{\sqrt{n}} \log \frac{W_\ell^n(Y_\ell^n|X^n)}{Q_{Y_\ell^n}(Y_\ell^n)} \le \frac{1}{\sqrt{n}} \log M_n + \frac{1}{\sqrt{n}} \log w_\ell -  \eta'  \right\} - e^{-\sqrt{n} \eta'} \label{eq:mixed_LB2b}
\end{align}
となる. 一方, \eqref{eq:mix_alpha_ineq4}による下界式において, $\gamma = Me^{-\sqrt{n} \eta'}$とおくと, 
\begin{align}
\varepsilon_n &\ge \sum_{\ell \in \Omega} w_\ell P_{X^nY_\ell^n} \left\{ \frac{1}{\sqrt{n}} \log \frac{W_\ell^n(Y_\ell^n|X^n)}{Q_{Y_\ell^n}(Y_\ell^n)} \le \frac{1}{\sqrt{n}} \log M_n -  \eta'  \right\} - e^{-\sqrt{n} \eta'} \label{eq:improved_mixed_LB1}
\end{align}
となりタイトになる. ここで, 有限長の$n$においては, $-\frac{1}{\sqrt{n}} \log w_\ell$の値が非常に大きくなる場合があることに注意しよう. この点は有限長解析において有用となることが期待される. 

\fi 

\section*{Acknowledgments}
The authors thank Prof. Te Sun Han for inspiring discussions.
This research was supported in part by MEXT under  Grant-in-Aid for Scientific Research (C) No.\ 25420357 and No.\ 26420371.

\end{document}